\documentclass[twocolumn]{aastex631}

\usepackage{natbib}
\usepackage{graphicx}
\usepackage{txfonts}
\usepackage{multirow}
\usepackage{xspace}
\usepackage{scalerel}
\usepackage[normalem]{ulem}
\usepackage{booktabs}
\usepackage{cleveref}
\usepackage{enumitem}

\shorttitle{The Instability Strip of LPVs}
\shortauthors{Trabucchi \& Pastorelli}
\begin{document}
\title{Self-Excited Pulsations and the Instability Strip of Long-Period Variables:\\the Transition from Small-Amplitude Red Giants to Semi-Regular Variables}
\author[0000-0002-1429-2388]{Michele Trabucchi}
\affiliation{Dipartimento di Fisica e Astronomia ``Galileo Galilei'', Universit\`a di Padova, Vicolo dell'Osservatorio 3, I-35122 Padova, Italy}
\affiliation{INAF-Osservatorio Astronomico di Padova, Vicolo dell’Osservatorio 5, I-35122 Padova, Italy}
\author[0000-0002-9300-7409]{Giada Pastorelli}
\affiliation{Dipartimento di Fisica e Astronomia ``Galileo Galilei'', Universit\`a di Padova, Vicolo dell'Osservatorio 3, I-35122 Padova, Italy}
\affiliation{INAF-Osservatorio Astronomico di Padova, Vicolo dell’Osservatorio 5, I-35122 Padova, Italy}

\received{2024 September 7}
\revised{2024 October 28}
\accepted{2024 November 8}
% \submitjournal{The Astrophysical Journal}

\begin{abstract}
We use one-dimensional hydrodynamic calculations combined with synthetic stellar population models of the Magellanic Clouds to study the onset of self-excited pulsation in luminous red giants. By comparing the results with Optical Gravitational Lensing Experiment observations in the period-luminosity (PL) diagram we are able to link the transition from small-amplitude red giants to semi-regular variables with a shift from stochastic driving to self-excited pulsations. This is consistent with previous studies relating this transition with an increase in mass-loss rate, dust formation, and the appearance of long secondary periods. The luminosity and effective temperature at the onset of pulsation are found to depend on metallicity, hydrogen content, and the adopted mixing length parameter. This confirms the role of partial hydrogen ionization in driving the pulsation, supporting the idea of a heat mechanism similar to that of classical pulsators. We examine the impact of turbulent viscosity, and find clear evidence that it must be adjusted according to the stellar chemical and physical parameters to fully match observations. In order to improve the predictive power of pulsation models, the turbulent viscosity and the temperature scale of pulsating red giants must be jointly calibrated. This is critical for model-based studies of the PL relations of evolved stars and to exploit their potential as distance and age indicators, in particular given the sensitivity of the onset of pulsation to the envelope composition. The grid of models is made publicly available with a companion interpolation routine.
\end{abstract}

\section{Introduction}
\label{sec:Introduction}

\begin{figure*}
    \centering
    \includegraphics[width=0.875\textwidth]{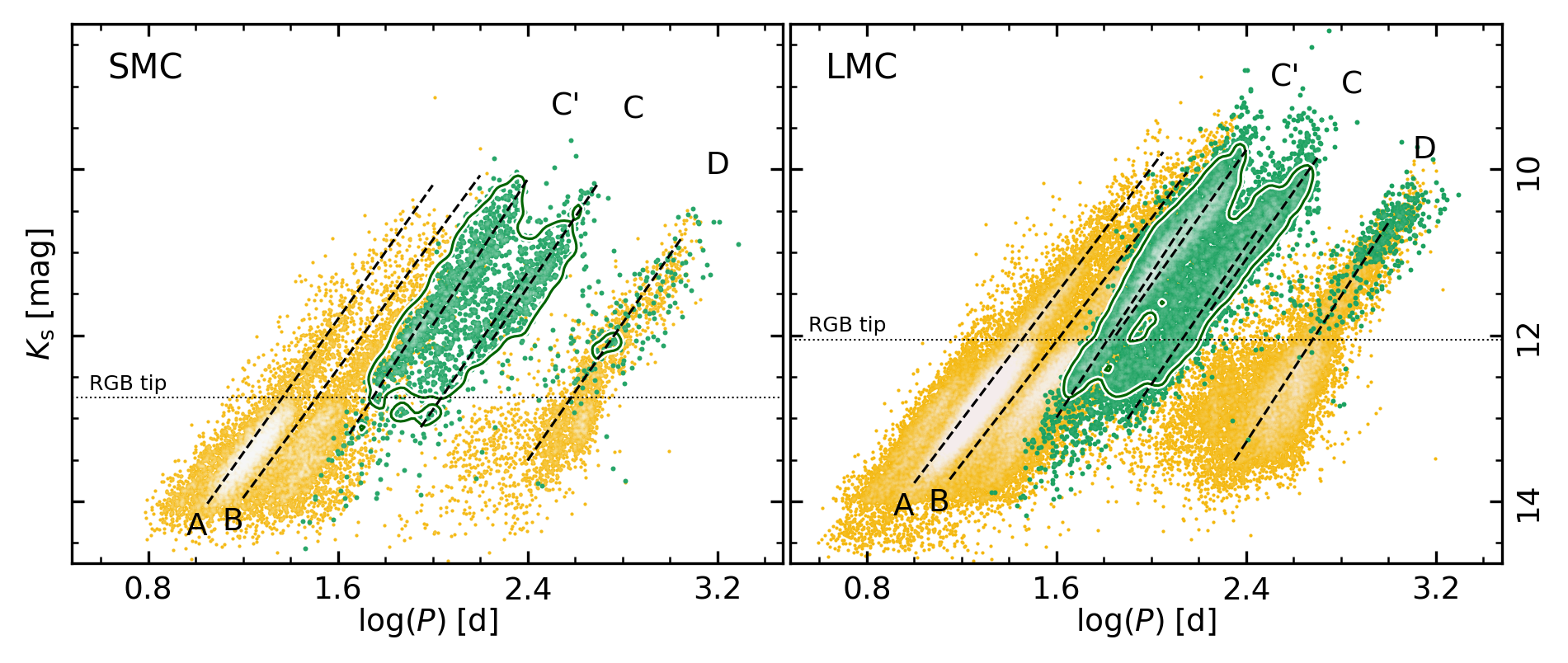}
    \caption{Period-luminosity diagrams of LPVs in the SMC (left-hand panel) and in the LMC (right-hand panel) from OGLE-III observations cross-matched with 2MASS photometry. Orange and green symbols mark OSARGs and SRVs, respectively (Miras are omitted). Dashed lines are the best fits to sequences A, B, C$^{\prime}$, C and D determined by \citet[][based on their Tables~1 and~2, and adopting the labels indicated in their Figures~2 and~4]{Soszynski_etal_2007}. For clarity, we do not distinguish between O- and C-rich SRVs in the scatter plot, but we show the best-fit lines to sequences C$^{\prime}$ and C for the respective populations following \citet{Soszynski_etal_2007}. The dotted lines indicate the approximate magnitude at the tip of the RGB. Density contour lines corresponding to 10\% of the maximum density of SRVs are overlaid.}
    \label{fig:OGLE3_KsPLDs}
\end{figure*}

Long-period variables (LPVs) represent a rather heterogeneous family of pulsating red stars. From the standpoint of stellar evolution, they consist primarily of red giant branch (RGB) and asymptotic giant branch (AGB) stars with low- to intermediate-mass ($1\lesssim M_{\rm i}\,/\,{\rm M}_{\odot}\lesssim8$) progenitors \citep{Herwig_2005,Wood_2015}, although more massive stars (up to $\approx25\,{\rm M}_{\odot}$) that become red supergiants (RSGs) during late core helium-burning stages may also exhibit LPV-like variability \citep[e.g.][]{Kiss_etal_2006}. They cover a vast range in terms of variability properties, with periods spanning from a few days to several years and photometric amplitudes up to about 10 mag in visual bands. As a result, observational limitations and research scopes have affected their classification, and a number of schemes have been proposed to distinguish between different subtypes of LPVs. Following a long-standing practice, stars with visual amplitudes larger than 2.5 mag are classified as Mira variables. These are among the most evolved and intrinsically brightest AGB stars, and have periods from a few months to a few years. Moving towards fainter and less evolved objects, one encounters shorter periods, smaller amplitudes, and a lesser degree of regularity, often related to the presence of multiple periods in the light curve. Some of these stars are classified as semi-regular (SR) variables, of which several subclasses exist \citep{Samus_etal_2017}. Other stars are often labeled as irregular variables or small-amplitude red variables/giants \citep[see, e.g.,][]{Eggen_1977}. An overview of the variability classification of red giants can be found in \citet{PulsatingStars_2015}.

Large-scale optical microlensing surveys initiated in the 1990s resulted in a wealth of photometric time series of variable stars in the Milky Way and the Magellanic Clouds (MCs), and have played a central role in the study of LPVs \citep{Wood_etal_1999,Wood_2000}. In particular, the Optical Gravitational Lensing Experiment \citep[OGLE;][]{Udalski_etal_1992} delivered a catalog of LPVs \citep{Soszynski_etal_2009,Soszynski_etal_2011,Soszynski_etal_2013,Iwanek_etal_2022} that represents a gold standard for the investigation of this type of stars. The catalog employs its own classification scheme based on the features of photometric time series in the Cousins $I$ band. Miras are identified by their peak-to-peak amplitude $\Delta I$ larger than $0.8$ mag, whereas the remaining stars are classified either as semi-regular variables (SRVs) or OGLE small-amplitude red giants (OSARGs) based on their period ratios and on the location of their multiple periods in the sequences and ridges of the period-luminosity (PL) diagram \citep{Soszynski_etal_2007}. There is evidence of an evolutionary sequence through these variability types \citep[e.g.][]{Soszynski_Wood_Udalski_2013,Trabucchi_etal_2019}, yet the relationships and differences between them are not entirely clear. OSARGs and SRVs populate distinct regions of the PL diagram (PLD; Fig.~\ref{fig:OGLE3_KsPLDs}), the former being found primarily on sequences A and B, whereas the latter are located along sequences C$^{\prime}$ and C. However, both sequences B and C$^{\prime}$ are associated with pulsation in the first-overtone mode \citep[1OM;][]{Yu_etal_2020}, thus the difference between OSARGs and SRVs is not only a matter of dominant pulsation modes. Even though the gap between sequences B and C$^{\prime}$ has been linked with the appearance of long secondary periods \citep[LSPs;][]{Trabucchi_etal_2017}, the difference in character between OSARGs and SRVs suggests a more fundamental physical transition, supported by observational evidence of increased mass-loss rates \citep{McDonald_Trabucchi_2019}. One possibility is that they differ in the driving mechanism of pulsation \citep{Soszynski_etal_2004}.

Stellar oscillations are very common in stars with an extended convective envelope, and they can often be attributed to acoustic standing waves \citep[e.g.][]{Asteroseismology_2010}. It is widely accepted that a solar-like stochastic convective driving is the cause of these oscillations in RGB stars at relatively low luminosity \citep[see, e.g., the review by][]{Chaplin_Miglio_2013}, while the situation is less clear for the more evolved LPVs. Early dynamical models of LPVs \citep[e.g.][]{Keeley_1970,Wood_1974}, later complemented with linear stability analysis \citep[e.g.][]{Fox_Wood_1982,Ostlie_Cox_1986}, largely revolved around the notion that LPVs result from a $\kappa$/$\gamma$-type heat mechanism similar to that of Cepheid variables and other classical pulsators, and highlighted the importance of the zone of partial hydrogen ionization in the driving mechanism. The interplay between convection and pulsation is a notoriously complex problem \citep[see, e.g., the review by][]{Houdek_Dupret_2015}, which created difficulties in early studies, somewhat mitigated in more recent iterations \citep[e.g.][]{Olivier_Wood_2005,Keller_Wood_2006}. Such works shed light on the role of convection in the driving of pulsation \citep[e.g.][]{Munteanu_etal_2005,Xiong_Deng_2013}, although a definitive identification of the driving mechanism remains elusive.

\citet{ChristensenDalsgaard_etal_2001} were among the first to propose solar-like stochastic driving as an alternative for the excitation of SR variables \citep[see also][]{Bedding_2003}. As outlined by \citet{Hartig_etal_2014}, and more recently by \citet{Yu_etal_2020}, the notion that LPVs might be solar-like oscillators has been extensively explored with the advent of modern asteroseismology, and has gathered substantial support both on observational and theoretical grounds \citep[e.g.][]{Dziembowski_Soszynski_2010,Takayama_etal_2013,Mosser_etal_2013,Stello_etal_2014,Yu_etal_2020,Cadmus_2024}. A recurring argument in favor of this hypothesis is that the global solar-like oscillation pattern can be extended from the RGB to brighter magnitudes, finding a good agreement with the PL relations of LPVs.

Yet the latter can be reproduced comparably well invoking stochastic driving \citep[e.g.][]{Wood_2015,Trabucchi_etal_2019}. Using linear, nonadiabatic pulsation models with a nonlocal and time-dependent treatment of convection, \citet{Xiong_etal_2018} concluded that red giants transition from stochastic to self-excited pulsation with increasing luminosity. This appears consistent with the transition in variability properties observed by \citet{Banyai_etal_2013}. \citet{Cunha_etal_2020} reach similar conclusions based on a phenomenological model, and identify both SR and Mira variables as classic-type pulsators. Moreover, self-excited hydrodynamic models are necessary to reconcile stellar evolution models with observations of Miras \citep{Trabucchi_etal_2021a}. Further evidence in favor of self-excitation comes from full ``star-in-a-box'' three-dimensional simulations of the entire convective envelope of AGB stars, which show a high degree of compatibility with variability observations \citep[e.g.][]{Freytag_etal_2017,Ahmad_etal_2023} and near-infrared interferometric measurements \citep{Paladini_etal_2018,RosalesGuzman_etal_2024}.

Whether they argue in favor of the solar-like or self-excited mechanism, most of these studies are based on linear models which are not able to capture the full complexity of the pulsation in LPVs. The early hydrodynamic models and the very recent three-dimensional simulations represent notable exceptions, but they cover a restricted space of parameters and are limited in scope. In this work, we compute a grid of one-dimensional hydrodynamic pulsation models to address systematically the onset of self-excited pulsation in bright red giants, and to investigate the transition from small-amplitude red giants to SRVs. This paper is structured as follows. In Section~\ref{sec:Methods} we describe our methods, and in Section~\ref{sec:Results} we present the results of our analysis. In that same section, we address the comparison of our predictions with observations. We discuss the implications of our findings in Section~\ref{sec:Discussion} and draw our conclusions in Section~\ref{sec:Conclusions}.

\section{Methods}
\label{sec:Methods}

\subsection{Grid Construction}
\label{ssec:GridConstruction}

Given the variety of physical and chemical properties characterizing AGB stars, a grid-based approach is necessary to investigate the onset of pulsation in LPVs. We follow the approach described in \citet{Trabucchi_etal_2021a,Trabucchi_etal_2019}, to which we refer the reader for more details on the numerical codes and physical assumptions. Briefly, we compute sequences of hydrostatic structures of red giant envelopes with progressively increasing luminosity, keeping fixed all physical and composition parameters other than the core mass, for which a core mass-luminosity relation is assumed. We stress that such luminosity sequences are not evolutionary tracks, as the adopted codes are decoupled from evolution models, and simplifications are made. For instance, the core region (including energy sources) is only treated as an inner boundary, and the chemical composition is assumed to be constant and homogeneous throughout the envelope.

For each hydrostatic model we perform a linear stability analysis against nonadiabatic radial oscillations, obtaining the linear period and growth rate of up to five pulsation modes, from the fundamental mode (FM) to the fourth-overtone mode. Finally, we probe the stability of the envelope against nonlinear radial pulsations by computing hydrodynamic one-dimensional time series that take the hydrostatic model as a starting point $t^0$. Each following model, say at time $t^n$, is computed after a time step $\delta t^n=t^n-t^{n-1}$, the value of which is assigned by an adaptive scheme. Briefly, a user-defined value (arbitrarily set to a few percent of the dominant linear period) is taken at first, and is then adapted so as to ensure that fractional changes in pressure and temperature between successive models remain smaller than about 10\% in all mass zones. As a result, the time step is not constant, but depends on the model properties. For the models explored here, the time step ranges from about an hour to about a day in presence of pulsation. Conversely, if the model is dynamically stable, the time step can become as long a year or more.

We construct a grid of 324 model sequences by varying the following parameters: total mass $M/{\rm M}_{\odot}\in\{1.0$, $2.6\}$, mixing length parameter $\alpha_{\rm ML}\in\{1.5$, $2.0$, $2.5\}$, hydrogen mass fraction $X\in\{0.6$, $0.7$, $0.8\}$, metallicity $Z\in\{0.002$, $0.006$, $0.010\}$, and turbulent viscosity parameter $\alpha_{\nu}\in\{0.00$, $0.01$, $0.02$, $0.05$, $0.10$, $0.20\}$. These values represent a subset of those adopted in \citet{Trabucchi_etal_2019,Trabucchi_etal_2021a}, and are adopted for consistency and for enabling direct comparison with previous results. While not all combinations are necessarily realistic, they provide the appropriate coverage for grid-based interpolation.
A relevant upgrade with respect to our previous works is the adoption of updated low-temperature Rosseland mean radiative opacities from \texttt{{\AE}SOPUS}~\texttt{2.0} \citep{Marigo_etal_2022}. We adopt the scaled-solar chemical composition by \citet{Caffau_etal_2011}, and in this work we limit our analysis to oxygen-rich composition. A more detailed study of the impact of carbon enrichment on pulsation will be addressed in a future work.

\subsection{Hydrodynamic Time Series}
\label{ssec:HydrodynamicTimeSeries}

For each envelope model we first compute a time series consisting of 2000 time steps. No initial perturbation is applied to the hydrostatic model, and instability (if any) is triggered by numerical noise. In the latter case, the integrated kinetic energy $E_{\rm k}$ increases rapidly within the first few hundred time steps, otherwise it decreases exponentially with time, providing a convenient way of detecting instability without having to compute an extended time series. We proceed this way for each envelope model with progressively increasing luminosity until the first unstable model is encountered.

\begin{deluxetable*}{ccccccccccDDDDD}
    \tablecaption{Values of the grid nodes (first five columns) and the corresponding model parameters at the first onset of nonlinear instability.}
    \label{tab:results}
    \tablewidth{0pt}
    \tablehead{\colhead{$X$} & \colhead{$Z$} & \colhead{$\alpha_{\rm ML}$} & \colhead{$M$} & \colhead{$\alpha_{\nu}$} & \colhead{$\log(L)^{\rm(a)}$} & \colhead{$T_{\rm eff}^{\rm(a)}$} & \colhead{$\log(\langle L\rangle)^{\rm(b)}$} & \colhead{$\langle T_{\rm eff}\rangle^{\rm(b)}$} & \colhead{$n_{\rm dom}^{\rm(c)}$} & \twocolhead{$P_{\rm dom}^{\rm(c)}$} & \twocolhead{$P_{\rm FM}^{\rm(d)}$} & \twocolhead{${\rm GR}_{\rm FM}^{\rm(d)}$} & \twocolhead{$P_{\rm 1OM}^{\rm(d)}$} & \twocolhead{${\rm GR}_{\rm 1OM}^{\rm(d)}$}
    \\
    \colhead{} & \colhead{} & \colhead{} & \colhead{$({\rm M}_{\odot})$} & \colhead{} & \colhead{$({\rm L}_{\odot})$} & \colhead{$({\rm K})$} & \colhead{$({\rm L}_{\odot})$} & \colhead{$({\rm K})$} & \colhead{} & \twocolhead{$({\rm days})$} & \twocolhead{$({\rm days})$} & \twocolhead{} & \twocolhead{$({\rm days})$} & \twocolhead{}
    }
    \decimals
    \startdata
    $0.6$ & $0.002$ & $1.5$ & $1.0$ &   $0.00$ & $3.250$ & $3819$ & $3.261$ & $3784$ &  $1$ &   $36.9$ &   $64.0$ &  $1.4351\cdot10^{-2}$ &  $36.1$ &  $6.9608\cdot10^{-2}$ \\
    $0.6$ & $0.002$ & $1.5$ & $1.0$ &   $0.01$ & $3.260$ & $3811$ & $3.270$ & $3776$ &  $1$ &   $37.6$ &   $66.0$ &  $1.2342\cdot10^{-2}$ &  $37.0$ &  $6.7823\cdot10^{-2}$ \\
    $0.6$ & $0.002$ & $1.5$ & $1.0$ &   $0.02$ & $3.300$ & $3778$ & $3.309$ & $3739$ &  $1$ &   $41.8$ &   $74.5$ &  $9.2930\cdot10^{-3}$ &  $41.0$ &  $7.1675\cdot10^{-2}$ \\
    $0.6$ & $0.002$ & $1.5$ & $1.0$ &   $0.05$ & $3.310$ & $3771$ & $3.317$ & $3732$ &  $1$ &   $43.0$ &   $76.8$ &  $2.9633\cdot10^{-3}$ &  $42.1$ &  $6.2310\cdot10^{-2}$ \\
    $0.6$ & $0.002$ & $1.5$ & $1.0$ &   $0.10$ & $3.365$ & $3726$ & $3.372$ & $3675$ &  $1$ &   $49.3$ &   $91.2$ & $-1.0838\cdot10^{-2}$ &  $48.4$ &  $5.3734\cdot10^{-2}$ \\
    $0.6$ & $0.002$ & $1.5$ & $1.0$ &   $0.20$ & $3.465$ & $3643$ & $3.465$ & $3624$ &  $1$ &   $64.7$ &  $126.3$ & $-4.2796\cdot10^{-2}$ &  $62.5$ &  $3.4600\cdot10^{-2}$ \\
    $0.6$ & $0.002$ & $1.5$ & $2.6$ &   $0.00$ & $4.100$ & $3588$ & $4.102$ & $3520$ &  $0$ &  $257.1$ &  $294.2$ &  $1.8391\cdot10^{-2}$ & $125.4$ &  $6.4487\cdot10^{-2}$ \\
    $0.6$ & $0.002$ & $1.5$ & $2.6$ &   $0.01$ & $4.115$ & $3575$ & $4.116$ & $3510$ &  $0$ &  $266.8$ &  $311.0$ &  $2.4189\cdot10^{-2}$ & $130.4$ &  $6.4513\cdot10^{-2}$ \\
    $0.6$ & $0.002$ & $1.5$ & $2.6$ &   $0.02$ & $4.135$ & $3559$ & $4.136$ & $3488$ &  $0$ &  $281.2$ &  $335.2$ &  $3.5492\cdot10^{-2}$ & $137.5$ &  $6.6657\cdot10^{-2}$ \\
    $0.6$ & $0.002$ & $1.5$ & $2.6$ &   $0.05$ & $4.155$ & $3542$ & $4.156$ & $3456$ &  $0$ &  $295.7$ &  $361.9$ &  $4.5907\cdot10^{-2}$ & $145.0$ &  $5.8806\cdot10^{-2}$ %\\
    \enddata
    \tablecomments{$^{\rm(a)}$ Luminosity and effective temperature from hydrostatic models. $^{\rm(b)}$ Time-averaged luminosity and effective temperature from hydrodynamic models. $^{\rm(c)}$ Variability properties from hydrodynamic models. $^{\rm(d)}$ Periods and growth rates of the FM and 1OM from linear stability analysis.\\Only a portion of this table is shown here to demonstrate its form and content. A machine-readable version of the full table is available.}
\end{deluxetable*}

At that point, we terminate the computation of the luminosity sequence, and extend the time series to $10^5$ time steps. We visually inspect the hydrodynamic time series of the unstable model to ensure that the pulsation attains the stable limit cycle. In a number of cases, the model may briefly pulsate in a higher-order mode before settling into the dominant mode. Albeit transient, this behavior can occasionally last for a significant fraction of the time series. In order to make sure that the model develops a truly dominant mode, we extend the time series iteratively until the pulsation maintains a fairly stable behavior for at least 50\% of the time series.

Once this criterion is met, we process the second half of the time series, thereby limiting the analysis to the stable limit cycle pulsation. We resample it to obtain even time steps, and we compute a periodogram to determine the dominant period, whose radial order is inferred by comparing it with the linear periods. Furthermore, we compute time-averaged values of the luminosity $\langle L\rangle$ and effective temperature $\langle T_{\rm eff}\rangle$. These are not necessarily the same as in the hydrostatic case due to feedback effects of the pulsation on the envelope structure \citep{YaAri_Tuchman_1996,Trabucchi_etal_2021a}. Time-averaged luminosities tend to be slightly brighter than hydrostatic ones, but are compatible within about 3\% for all models. On the other hand, the effect on temperature is more pronounced, with hydrodynamic models being often cooler than hydrostatic ones by up to 6\% in the most extreme cases (corresponding to roughly $200\,{\rm K}$). We caution the reader that these trends are specific to the limited space of parameters explored here, and may not hold in general. We provide both the hydrostatic and time-averaged values, but we note that neither is preferable to the other in an absolute sense. Indeed, the latter likely represents a better depiction of the physical behavior of pulsating red giants, hence we adopt them for the purpose of directly comparing the models' grid with observations (Sect.~\ref{ssec:ComparisonWithObservations}). We do so by computing the absolute (mean) brightness in the $K_{\rm s}$ band from $\langle L\rangle$ and $\langle T_{\rm eff}\rangle$, using bolometric corrections derived from the synthetic spectra of \citet{Aringer_etal_2016}. Conversely, if one aims to combine the grid of pulsation models with evolutionary tracks or synthetic stellar population models (as presented in Sect.~\ref{ssec:CombinationWithStellarPopulationModels}), which are based on hydrostatic models, the hydrostatic values should be used for consistency.

\section{Results}
\label{sec:Results}

\begin{figure*}
    \centering
    \includegraphics[width=0.495\textwidth]{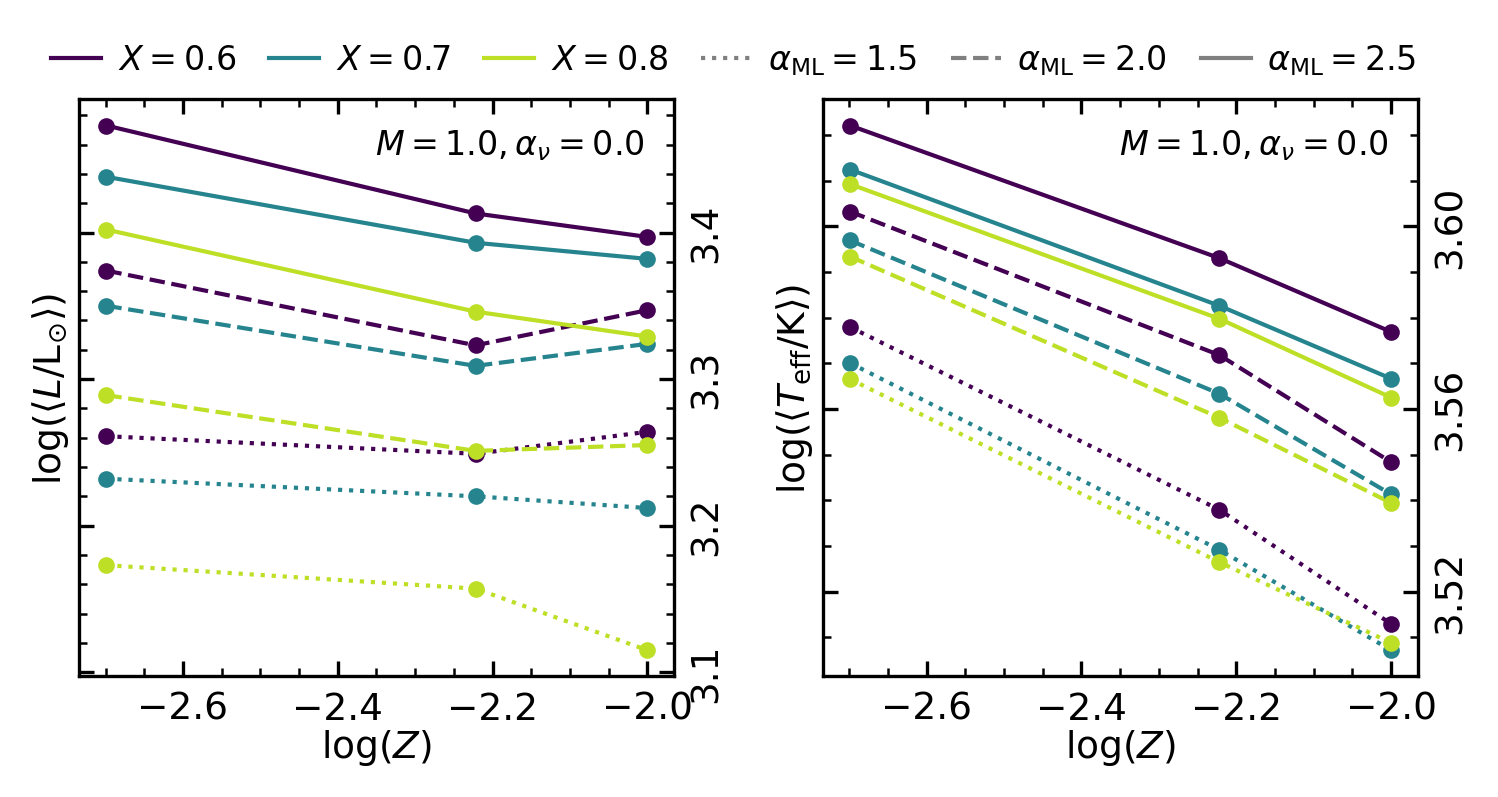}
    \includegraphics[width=0.495\textwidth]{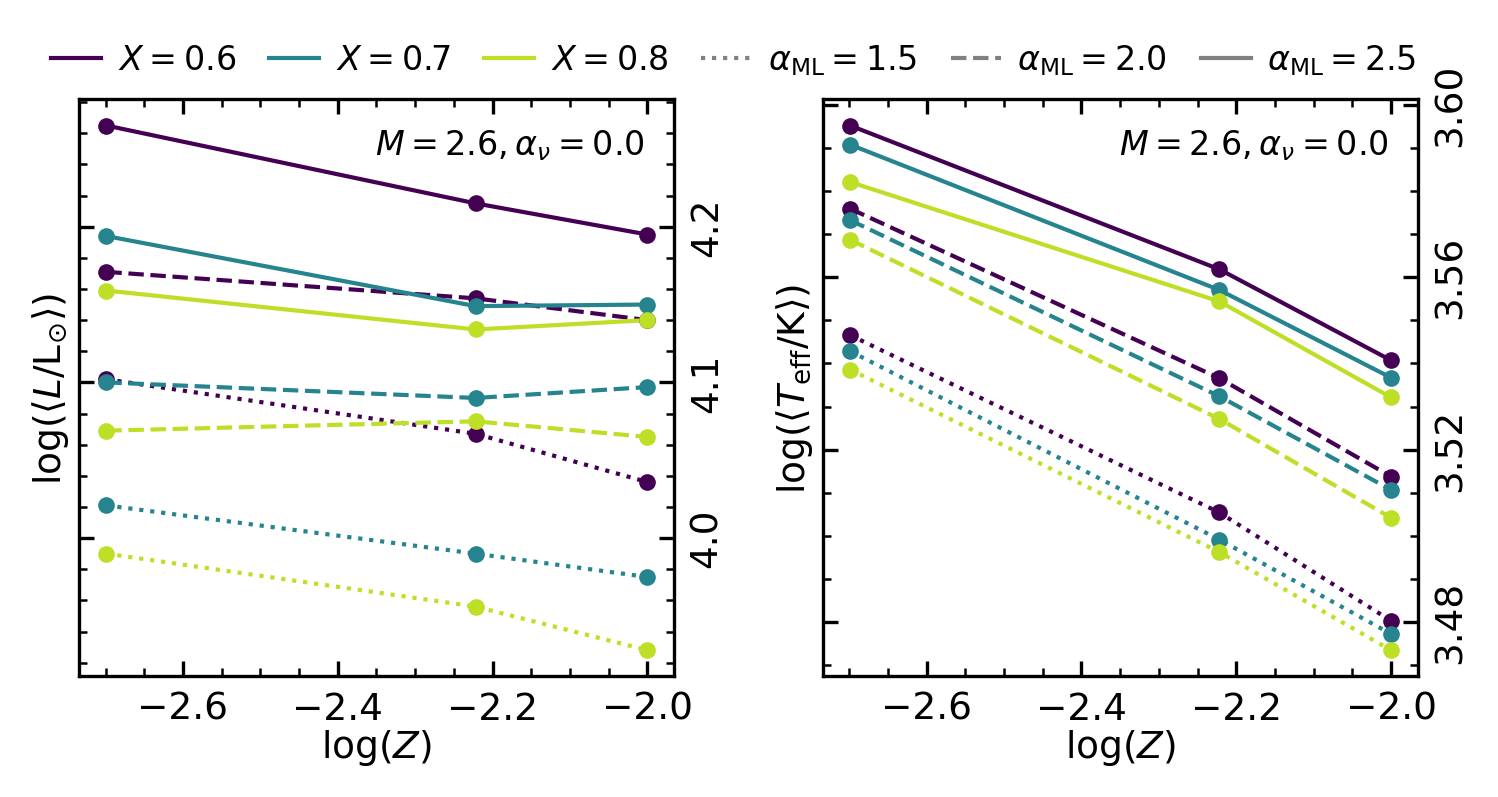}
    \includegraphics[width=0.495\textwidth]{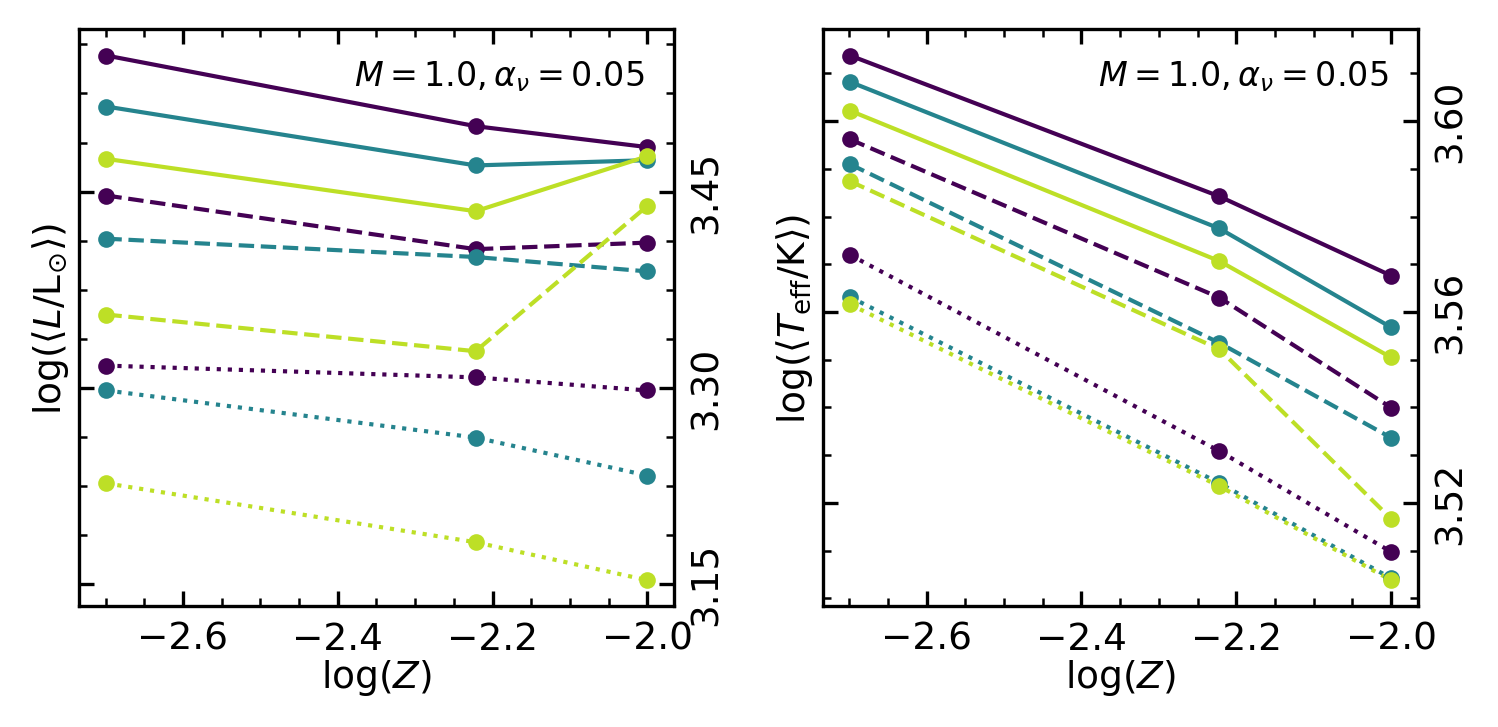}
    \includegraphics[width=0.495\textwidth]{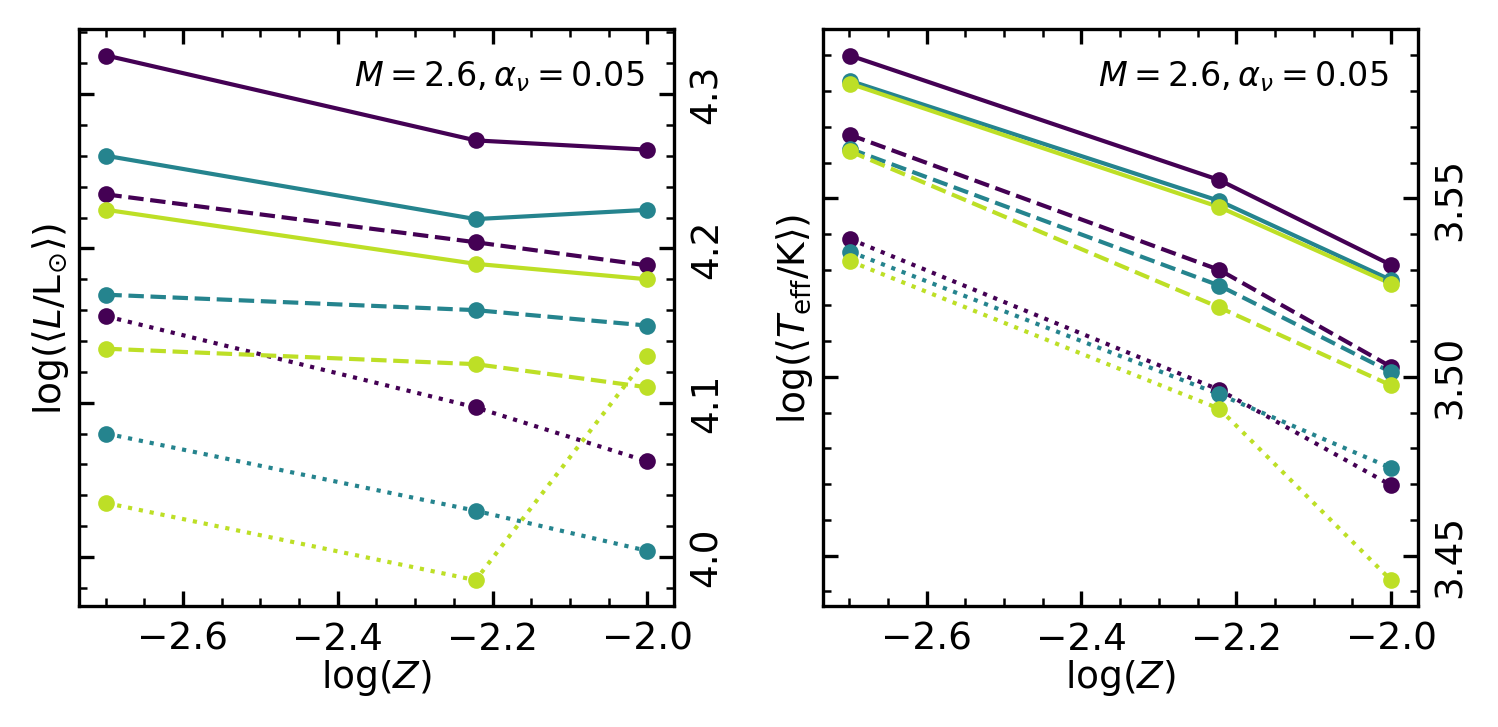}
    \includegraphics[width=0.495\textwidth]{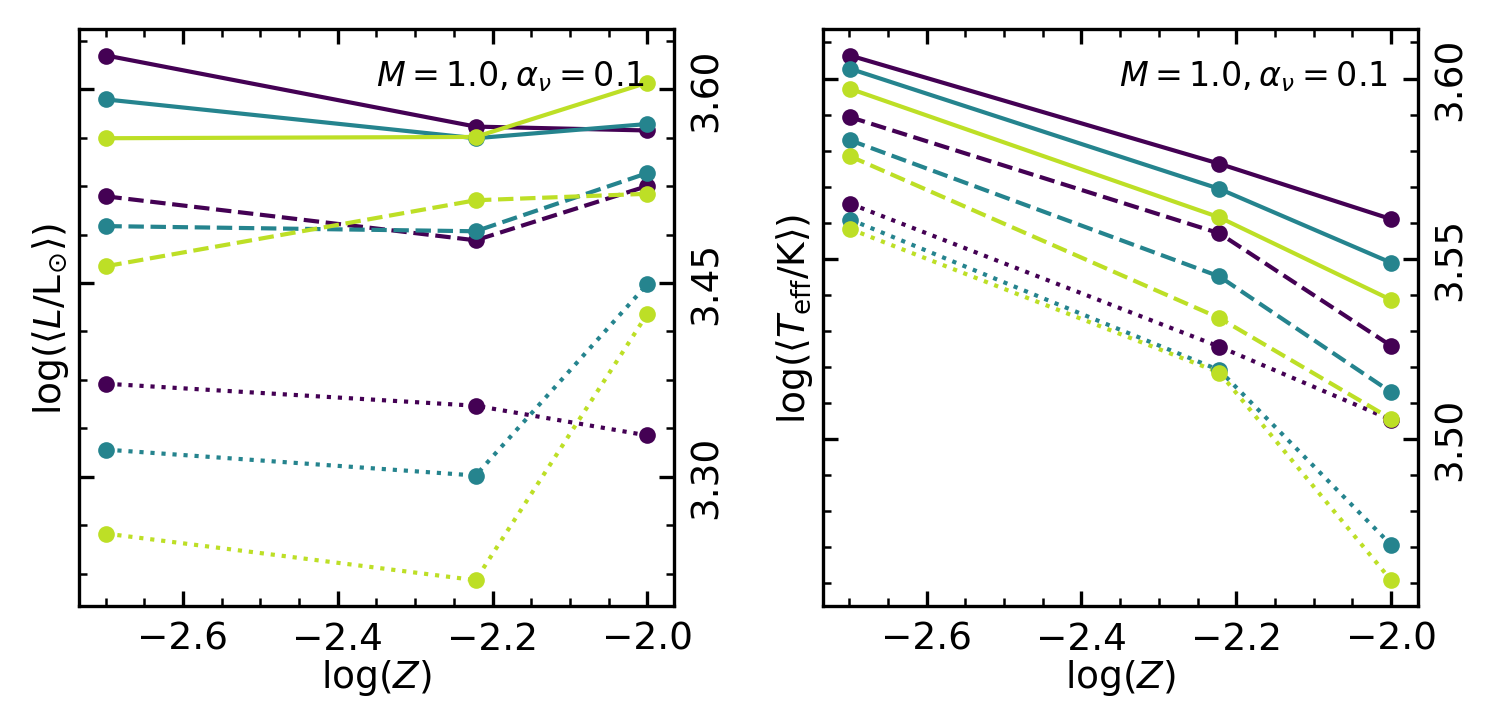}
    \includegraphics[width=0.495\textwidth]{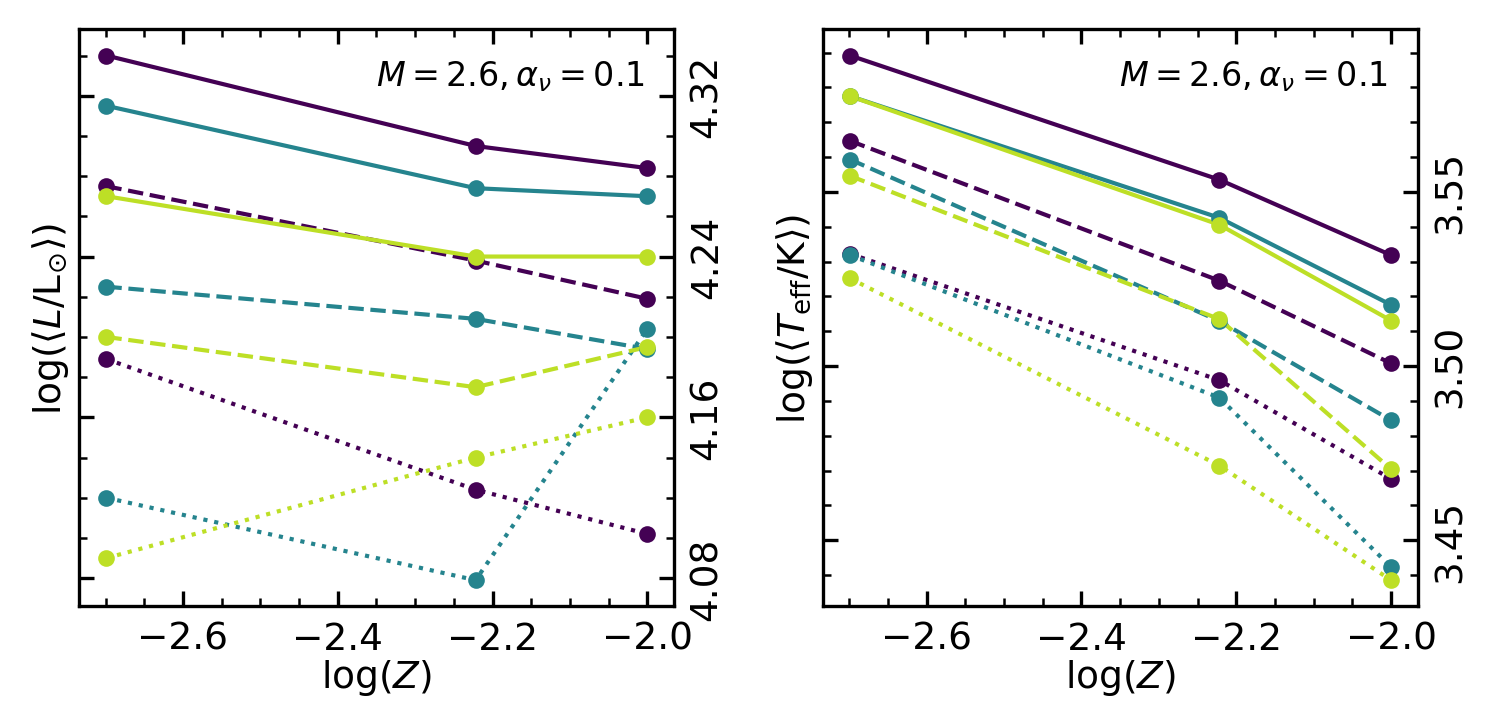}
    \includegraphics[width=0.495\textwidth]{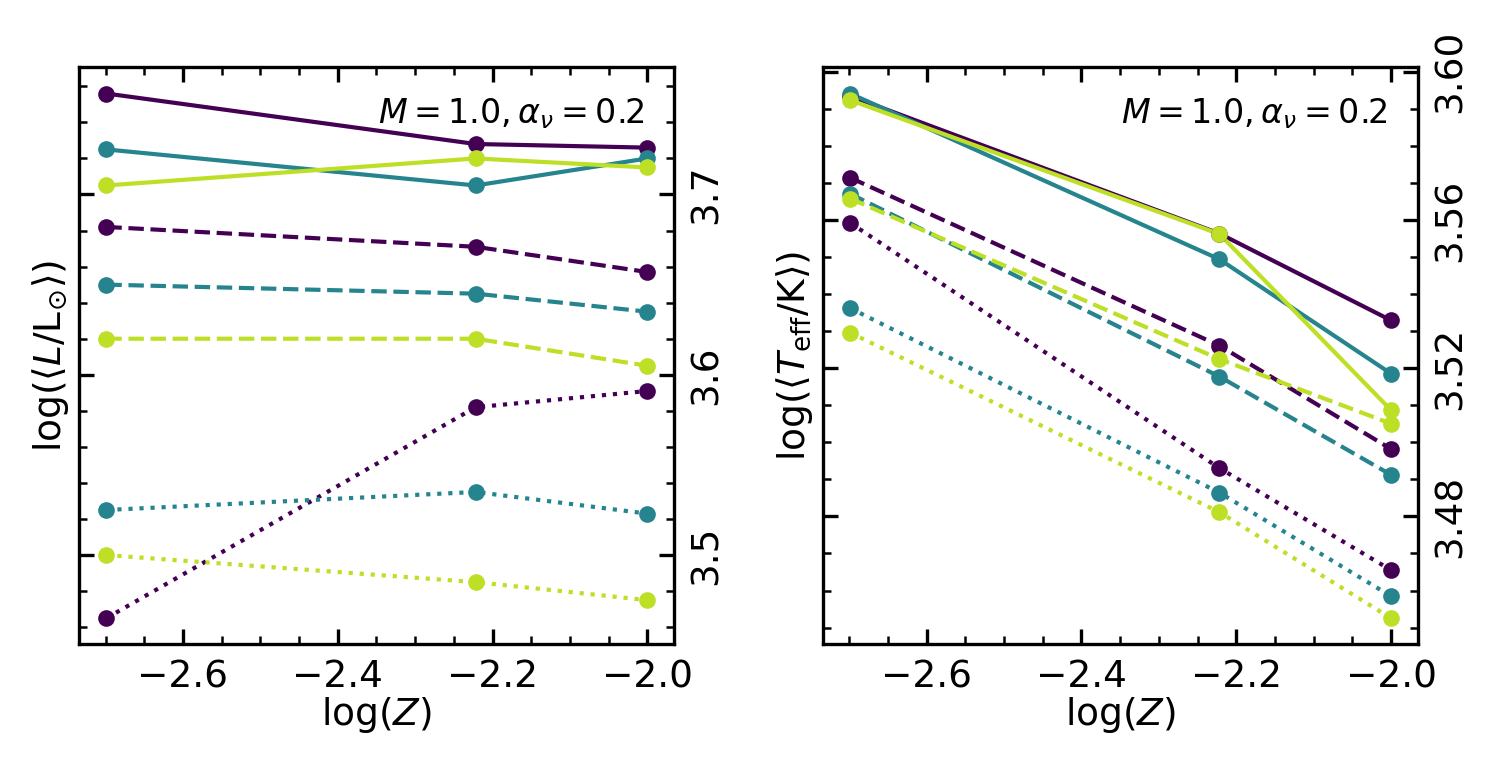}
    \includegraphics[width=0.495\textwidth]{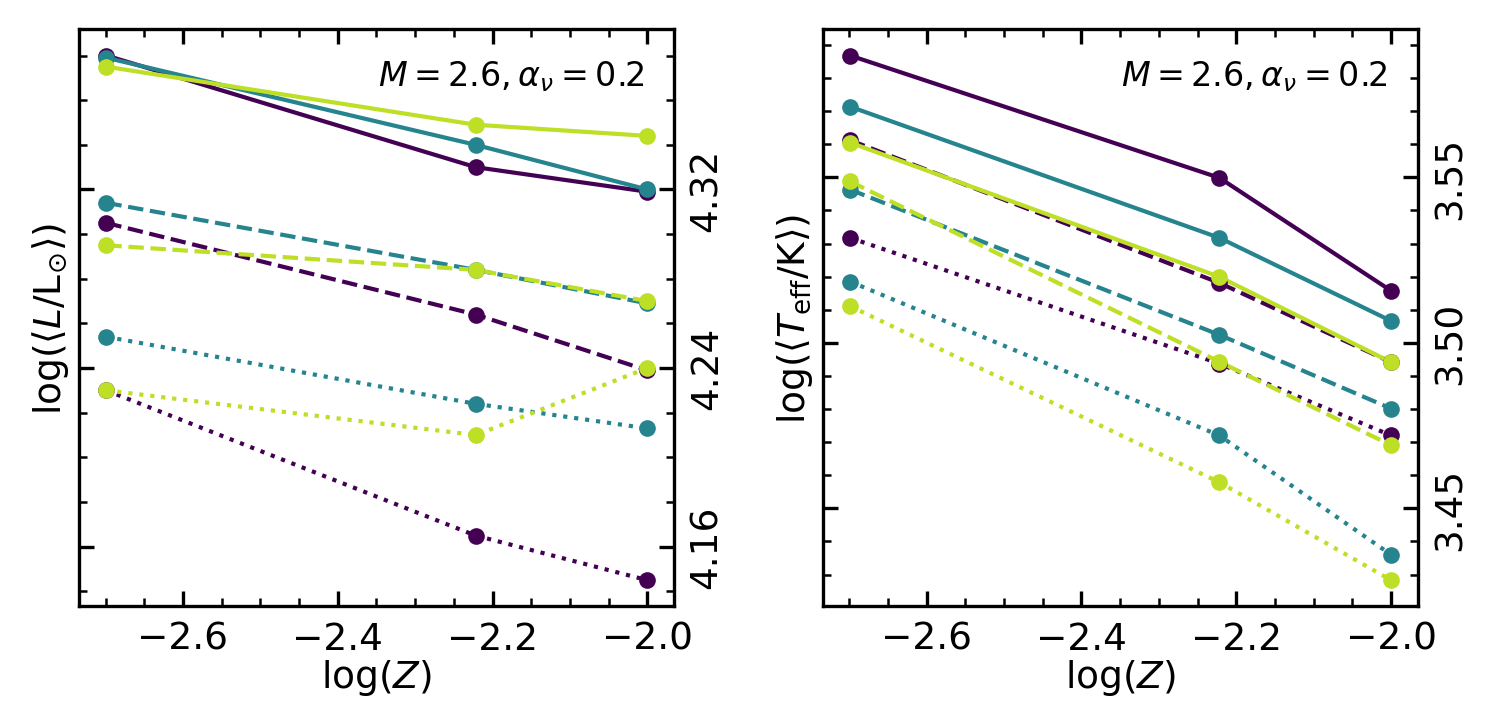}
    \caption{Model properties at the onset of instability as a function of the grid parameters. The panels in the first and second (third and fourth) columns from the left display luminosity and effective temperature as a function of luminosity for the $1.0\,{\rm M}_{\odot}$ ($2.6\,{\rm M}_{\odot}$) models.
    %The panels in the rightmost column show the location of the left edge of the instability strip (see text).
    The different rows correspond to different values of the turbulent viscosity parameter $\alpha_{\nu}$. Values of hydrogen abundances are indicated by line colors (blue: $X=0.60$, green: $X=0.70$, red: $X=0.80$), and values of the mixing length parameter by the line styles (dotted: $\alpha_{\rm ML}=1.5$, dashed: $\alpha_{\rm ML}=2.0$, solid: $\alpha_{\rm ML}=2.5$). Filled and empty symbols in the rightmost-column panels correspond to $1.0\,{\rm M}_{\odot}$ and $2.6\,{\rm M}_{\odot}$ models, respectively.}
    \label{fig:ZLT}
\end{figure*}

The results of our calculations for each grid node (combination of $X$, $Z$, $\alpha_{\rm ML}$, $M$, and $\alpha_{\nu}$) are collected in Table~\ref{tab:results}, which provides the hydrostatic and time-averaged values of $L$ and $T_{\rm eff}$ at the onset of self-excited pulsations, the dominant period $P_{\rm dom}$ and its radial order $n_{\rm dom}$, and the corresponding periods $P_n$ and growth rates ${\rm GR}_n$ of the FM ($n=0$) and 1OM ($n=1$) based on linear predictions. An overview of the results is provided in Fig.~\ref{fig:ZLT}, to which we refer to illustrate the general trends emerging from our models.

To begin with, we note that each one of our models becomes unstable either in the FM or in the 1OM. The analysis of multiperiodicity in nonlinear pulsation models is beyond the scope of this paper and will be addressed in a future work, yet we performed a visual inspection of the periodograms for a limited sample of models, and note that higher-order overtones may be temporarily excited, but they are never dominant. The 1OM is more frequently dominant in the $1.0\,{\rm M}_{\odot}$ models, whereas the FM tends to be favored in the $2.6\,{\rm M}_{\odot}$ ones. This is due to the latter having larger radii at the onset of instability, and indeed they display systematically larger values of $\langle L\rangle$ and cooled $\langle T_{\rm eff}\rangle$. In particular, the lower-mass models become unstable at about $1300$--$2950\,{\rm L}_{\odot}$ in the absence of turbulent viscosity ($\alpha_{\nu}=0$) and around $3000$--$5750\,{\rm L}_{\odot}$ if $\alpha_{\nu}=0.20$. For the more massive models these luminosity ranges are, respectively, around $8500$--$18600\,{\rm L}_{\odot}$ and $14000$--$24000\,{\rm L}_{\odot}$. The effective temperature at the onset of instability is in the range $2700\lesssim T_{\rm eff}\,/\,{\rm K}\lesssim4000$, a range that is little affected by either $\alpha_{\nu}$ or $M$.

Focusing on composition effects, we find that both $\langle L\rangle$ and $\langle T_{\rm eff}\rangle$ tend to decrease with increasing metallicity and hydrogen content, and with decreasing efficiency of superadiabatic convection (lower mixing length parameter). This is especially the case when no turbulent viscosity is applied. However, the picture becomes less regular when $\alpha_{\nu}$ is increased, in which case we find trends that are somewhat erratic. For instance, toward higher metallicity and for small $\alpha_{\rm ML}$ the luminosity increases with $Z$ and $X$ rather than decreasing. Conversely, temperature can drop more steeply with increasing $Z$, and these trends can be dissimilar for models with different masses. Overall, we find that the models depend in a complex fashion on the grid parameters, which is difficult to describe with a simple analytic relation.

\subsection{The Instability Strip}
\label{ssec:TheInstabilityStrip}

\begin{figure*}
    \centering
    \includegraphics[width=0.96\textwidth]{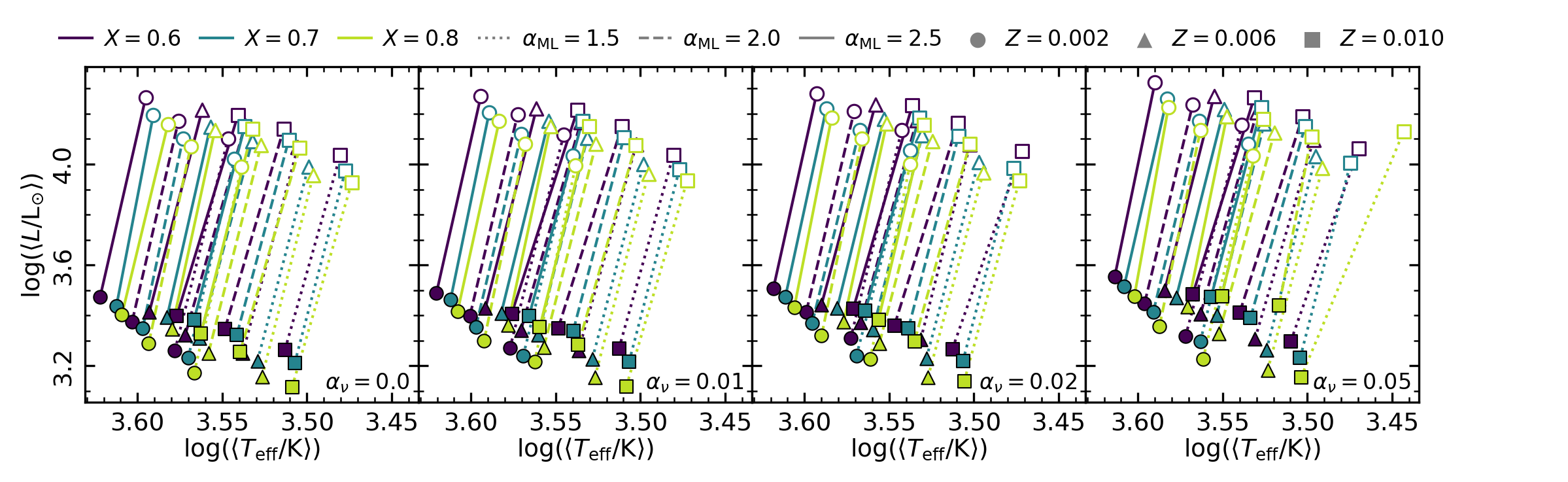}
    \includegraphics[width=0.96\textwidth]{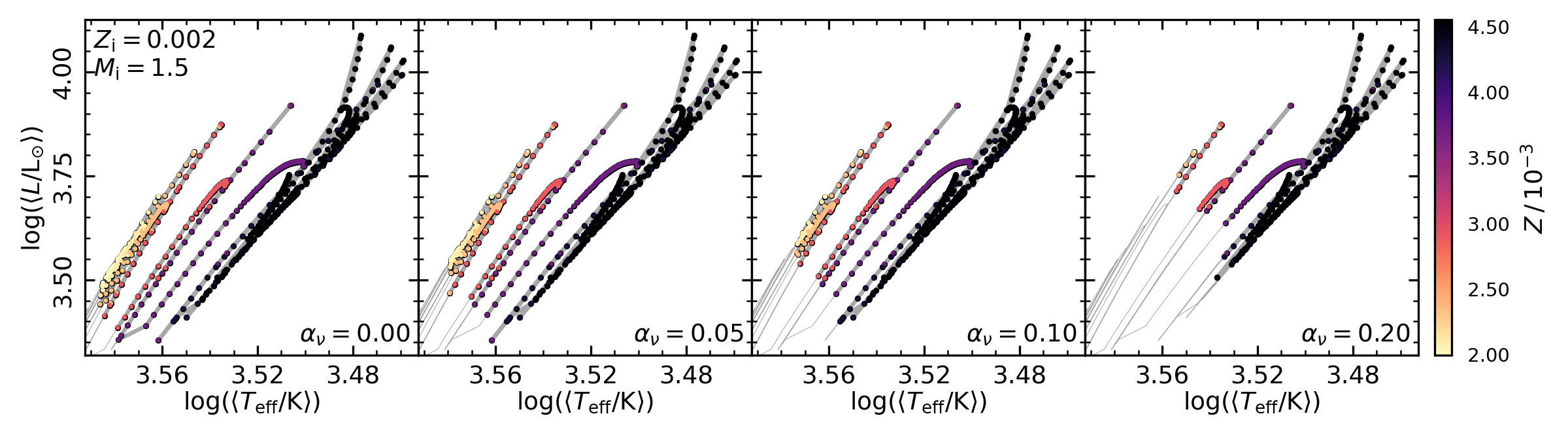}
    \includegraphics[width=0.96\textwidth]{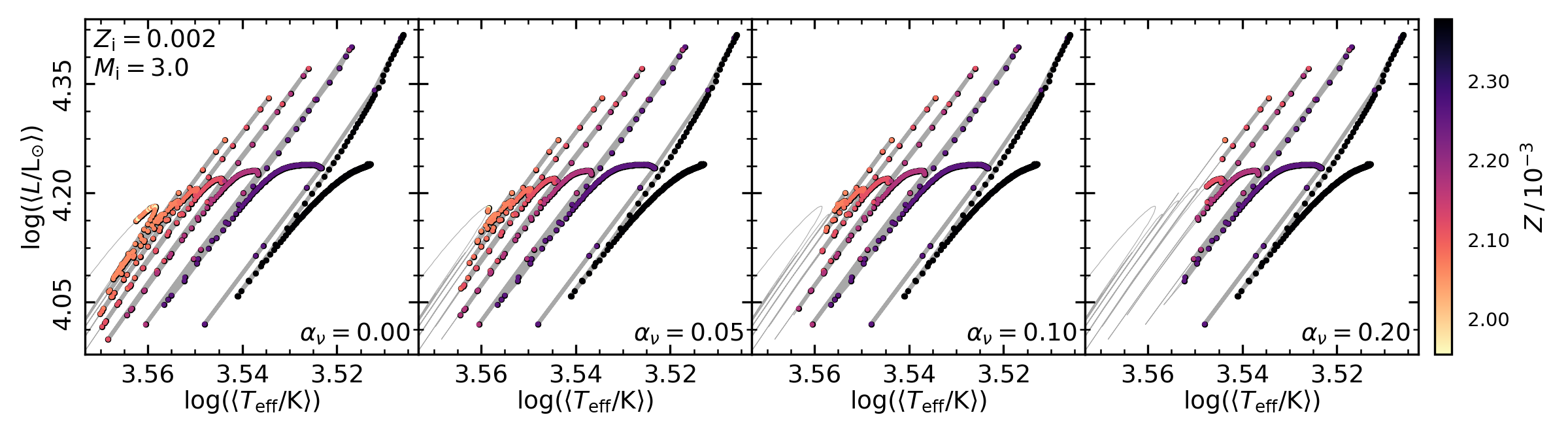}
    \includegraphics[width=0.96\textwidth]{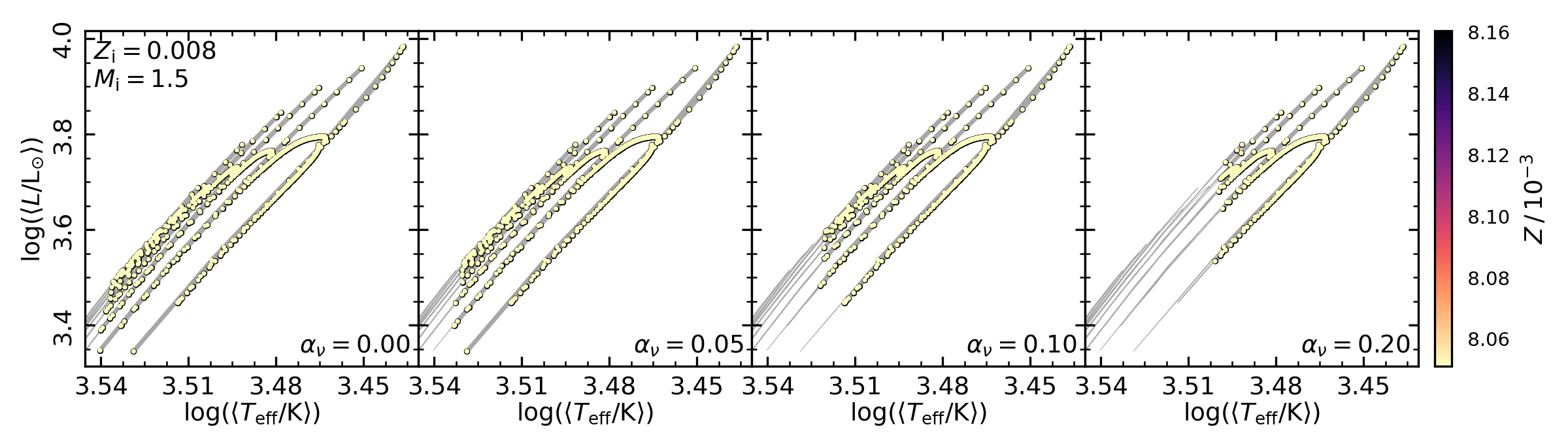}
    \caption{Top row: location of the blue edge of the instability strip in the HRD, with values of metallicity and mass indicated by symbols (circles: $Z=0.002$, triangles: $Z=0.006$, squares: $Z=0.010$; filled and empty symbols correspond to $1.0$ and $2.6\,{\rm M}_{\odot}$ models, respectively), while line colors and styles have the same meaning as in Fig.~\ref{fig:ZLT}. Other rows: pulsational instability along selected TP-AGB evolutionary tracks with initial metallicity and mass (from top to bottom) $Z_{\rm i}=0.002,\,M_{\rm i}=1.5$, $Z_{\rm i}=0.002,\,M_{\rm i}=3.0$, and $Z_{\rm i}=0.008,\,M_{\rm i}=1.5$. Panels on columns from left to right correspond to increasing values of $\alpha_{\nu}=0.00$, $0.05$, $0.10$, and $0.20$. The thin and thick portions of each track indicate the regimes in which pulsation is stable and unstable, respectively. Symbols along the latter portion are color-coded according to the current metallicity $Z$. For visual clarity, we omit late portions of the tracks, which bend toward higher $T_{\rm eff}$ approaching post-AGB stages.}
    \label{fig:evo}
\end{figure*}

By analogy with Classical Cepheids (CCs), it is instructive to examine these trends in terms of an instability strip (IS) in the Hertzsprung-Russell diagram (HRD). As we are studying the onset of instability, we focus on the LPVs' equivalent of the blue edge of the CCs' IS. This is shown in the panels in the top row of Fig.~\ref{fig:evo}. Regardless of the model composition and input physics, the edge of the IS is almost vertical, with a slight tendency to become even steeper at higher $\alpha_{\nu}$ due to the stronger impact of this parameter at lower masses.

As a result of the previously illustrated trends, the blue edge of the IS tends to become redder with increasing hydrogen and metal content, and lower $\alpha_{\rm ML}$. In this, it shows a behavior comparable, for esxample, with that of CCs \citep[][and references therein]{DeSomma_etal_2022}. However, the dependence of the IS on chemical composition has deeper implications in the case of LPVs compared to classical pulsators. Indeed, the IS predicted by our models often lies well within the space of stellar parameters typical of the AGB phase, so that stars can cross its blue edge back and forth multiple times if they undergo thermal pulses (TPs). If the composition is altered by third dredge-up (3DU) events, or the total mass is reduced by stellar winds, the blue edge of the IS will be different at each crossing. In particular, the luminosity corresponding to the onset of instability will be lower as a result of both processes. To investigate this effect we adopt thermally pulsing asymptotic giant branch (TP-AGB) evolutionary tracks computed with the \texttt{COLIBRI} code \citep{Marigo_etal_2013,Marigo_etal_2017}, accurately calibrated against observations in the MCs \citep{Pastorelli_etal_2019,Pastorelli_etal_2020}. We follow the grid-based interpolation approach described in \citep{Trabucchi_etal_2019} to determine, at each step along the track, the values of $L$ and $T_{\rm eff}$ at the onset of instability\footnote{
    The interpolation code is available online via\dataset[DOI: 10.5281/zenodo.14002294]{https://doi.org/10.5281/zenodo.14002294} \citep{14002294}.
}. For the purpose of interpolation, we employ the hydrostatic values of $L$ and $T_{\rm eff}$ in our grid rather than the time-averaged values, for consistency with the evolutionary models, which are computed in the hydrostatic approximation. As the difference is generally small, this choice does not affect our results. We repeat this interpolation scheme independently for each value of $\alpha_{\nu}$.

\begin{figure*}
    \centering
    \includegraphics[width=0.33\textwidth]{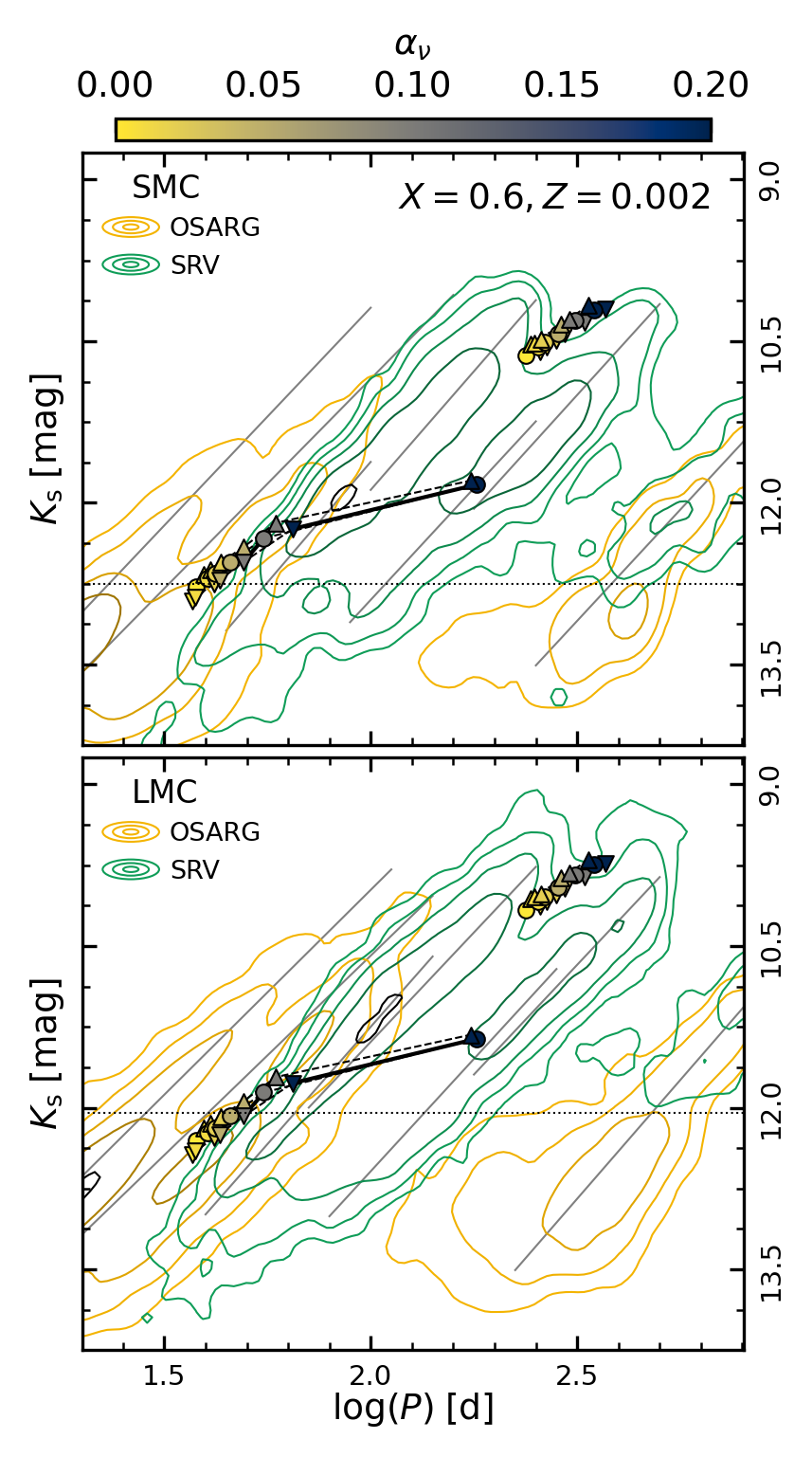}
    \includegraphics[width=0.33\textwidth]{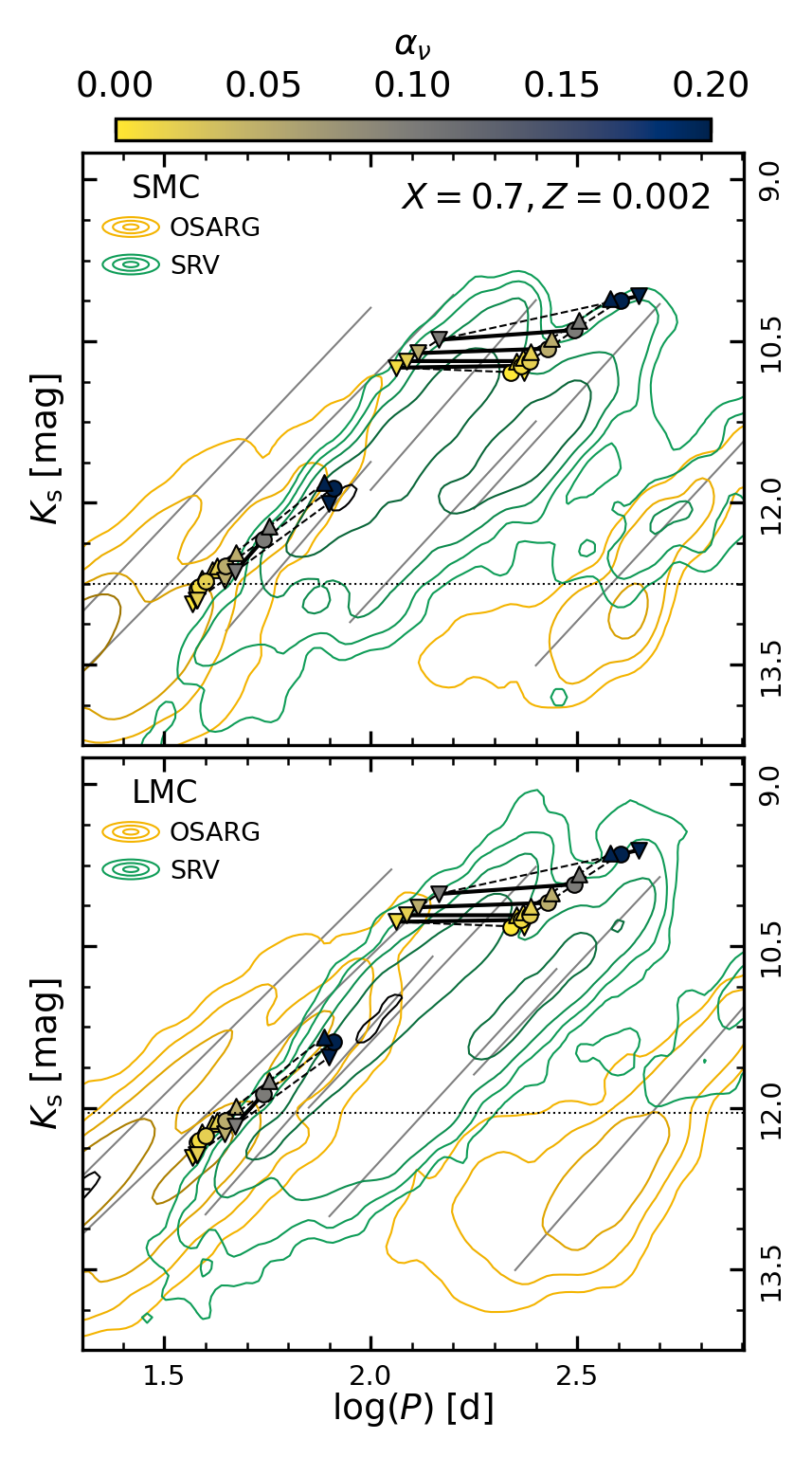}
    \includegraphics[width=0.33\textwidth]{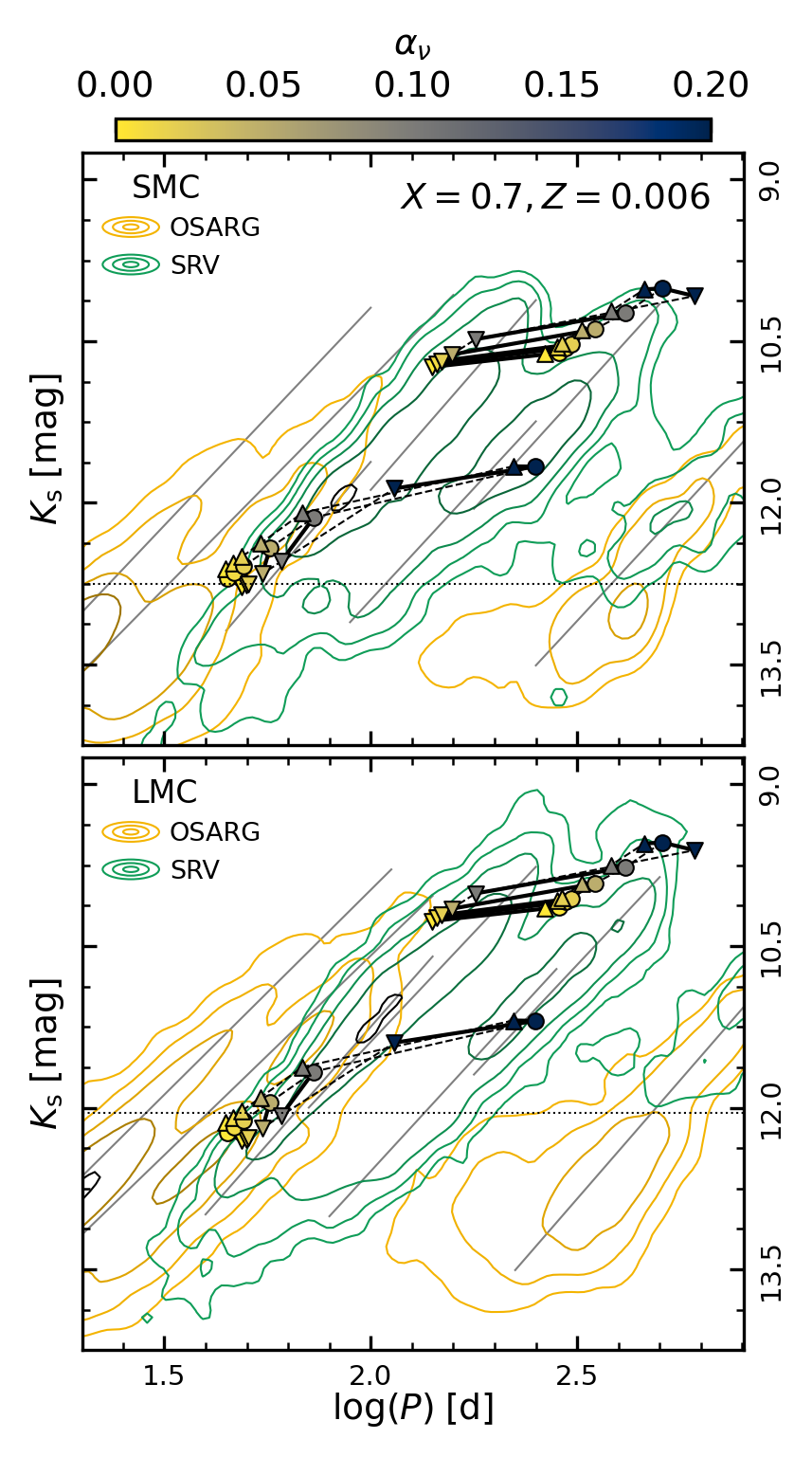}
    \caption{Example models overlaid to $K_{\rm s}$-band PL relations observed in the SMC (top panels) and LMC (bottom panels). Orange (green) lines indicate isodensity contours of OSARGs (SRVs), as in Fig.~\ref{fig:OGLE3_KsPLDs}. No distinction is made between O- and C-rich SRVs. Slanted gray lines and horizontal dotted lines have the same meaning as in Fig.~\ref{fig:OGLE3_KsPLDs}. Models are indicated by symbols, color-coded by their $\alpha_{\nu}$ and with different shapes for different $\alpha_{\rm ML}$ (down-pointing triangles: $\alpha_{\rm ML}=1.5$; circles: $\alpha_{\rm ML}=2.0$; up-pointing triangles: $\alpha_{\rm ML}=2.5$). Models with the same $\alpha_{\nu}$ (same $\alpha_{\rm ML}$) are connected by solid (dashed) black lines. The $X$, $Z$ values, fixed for models in a given column, are indicated in the top-right corner. We note that these models are selected to convey the general trends of the onset of pulsation in the PLD as a function of model parameter and composition rather than as representative of the LPVs in the MCs.}
    \label{fig:PLD_grid}
\end{figure*}

Figure~\ref{fig:evo} illustrates the unstable portions of three evolutionary tracks, selected to highlight specific aspects of the interplay between instability, composition, and mass loss. In the first case (second row of panels from the top), we show a track with initial metallicity $Z_{\rm i}=0.002$ and initial mass $M_{\rm i}=1.5\,{\rm M}_{\odot}$. Its surface metallicity increases significantly at each TP while its $T_{\rm eff}$ decreases, hence the star remains unstable to pulsation down to lower luminosities at each successive TP. This effect is enhanced by mass loss, which reduces the current mass of the model. With increasing turbulent viscosity we find a similar behavior, except it is pushed toward higher luminosities and cooler effective temperatures.

In the bottom half of Fig.~\ref{fig:evo} we display the TP-AGB evolution of models with $Z_{\rm i}=0.002,\,M_{\rm i}=3.0\,{\rm M}_{\odot}$ and $Z_{\rm i}=0.008,\,M_{\rm i}=1.5\,{\rm M}_{\odot}$, respectively. The 3DU is weak in the former and absent in the latter, so the IS is shaped primarily by mass loss. If we consider the hottest and faintest points reached during each TP while maintaining instability, we find them to describe a nearly vertical line in the HRD which is essentially the blue edge of the IS at fixed chemical composition, in striking contrast to the behavior of the first case.

\subsection{Comparison with Observations}
\label{ssec:ComparisonWithObservations}

For the purpose of comparing models with observations, we employ the catalogs of LPVs in the Large and Small Magellanic Clouds (LMC and SMC) from OGLE-III observations \citep{Soszynski_etal_2009,Soszynski_etal_2011}, and we include NIR photometry from the Two Micron All Sky Survey \citep[2MASS;][]{Skrutskie_etal_2006} by cross-matching the two datasets within 1$^{\prime\prime}$. The PLD of OSARGs and SRVs is shown in Fig.~\ref{fig:OGLE3_KsPLDs} as a scatter plot, a form of visualization that is not necessarily ideal for comparing with observations. Instead, we prefer to use isodensity contours to represent the distribution of observed primary periods of LPVs, as shown in Fig.~\ref{fig:PLD_grid}. To guide the eye, we also indicate the best-fit lines to sequences A, B, C$^{\prime}$ and C by \citet{Soszynski_etal_2007}.

The panels in the left column of Fig.~\ref{fig:PLD_grid} show the luminosity and dominant period at the onset of instability for $X=0.6$ and $Z=0.002$. To compute the apparent $K_{\rm s}$-band brightness at the distance of the MCs, we take the distance moduli of $\mu_{\rm SMC}=18.96$ mag and $\mu_{\rm LMC}=18.49$ mag recommended by \citet{deGrijs_etal_2017}. A striking feature of the model predictions is that they lie very close to the short-period edges of the observed distributions corresponding to the C$^{\prime}$ sequence of 1OM-dominated SRVs (for the $1.0\,{\rm M}_{\odot}$ models) and to the C sequence of FM-dominated SRVs (for the $2.6\,{\rm M}_{\odot}$ models). Interestingly, the models roughly trace the border of the sequences as $\alpha_{\nu}$ is varied. In some cases, adopting different values for the free parameters $\alpha_{\nu}$ and $\alpha_{\rm ML}$ results in the models shifting from one sequence to the other, sticking to the short-period edge. This effect is more evident if $X$ is increased to $0.7$ (central column of Fig.~\ref{fig:PLD_grid}), or if $Z$ is increased to $0.006$ (right column of Fig.~\ref{fig:PLD_grid}). Other combinations of $X$ and $Z$ show similar trends.

It is worth observing that, on their bright ends (above $K_{\rm s}\simeq11.5$ mag in the SMC and $K_{\rm s}\simeq10.5$ mag in the LMC), the observed SRV sequences are populated almost exclusively by C-rich stars. It is therefore interesting that the models display such a good match even though we have only considered O-rich compositions. We also note that in the vicinity of the tip of the RGB, where the SRVs become more scarce compared with OSARGs, the identification of the theoretical onset of instability with the edge of sequence C$^{\prime}$ is not as straightforward as it is at brighter magnitudes, especially at relatively low metallicity. Both these aspects will be further addressed in Sect.~\ref{sec:Discussion}.

\begin{figure*}
    \centering
    \includegraphics[width=0.33\textwidth]{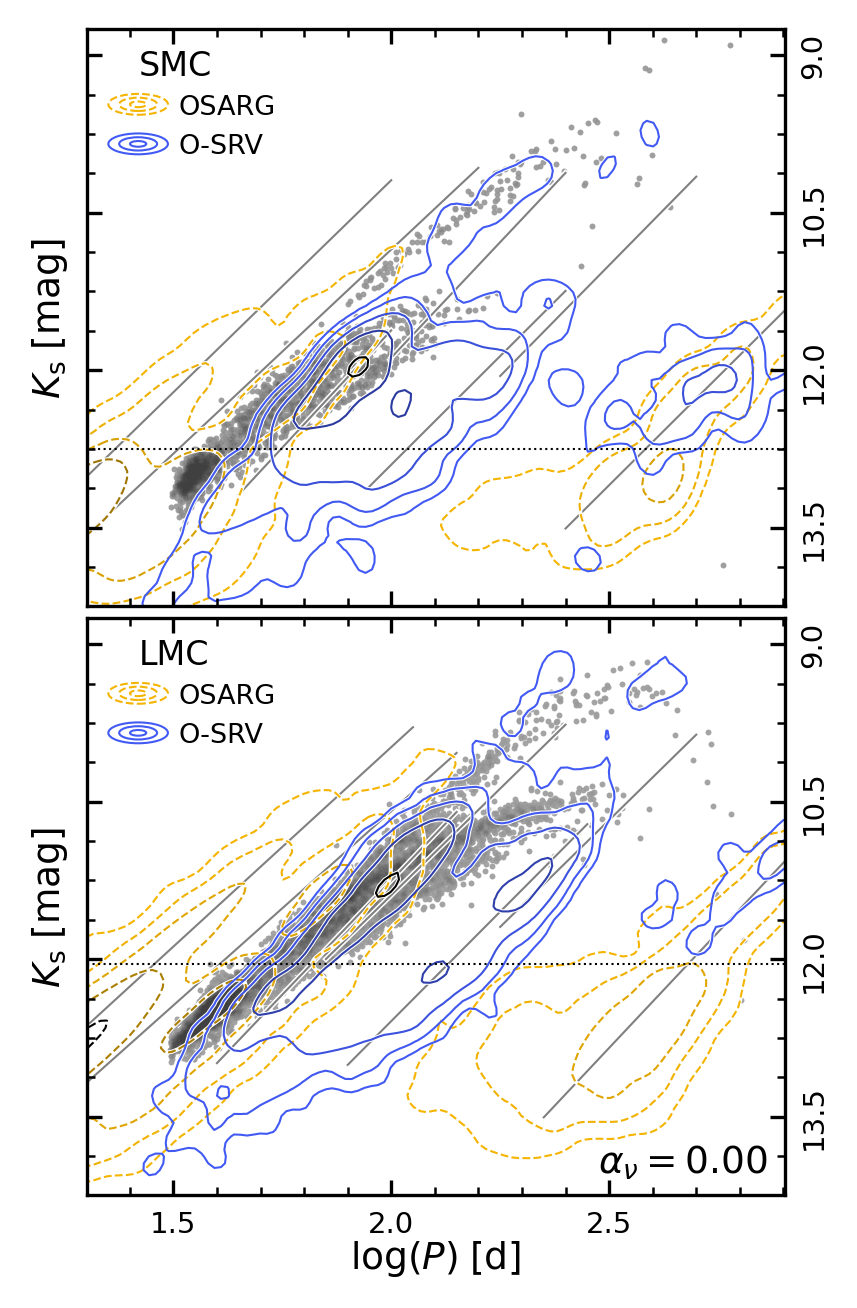}
    \includegraphics[width=0.33\textwidth]{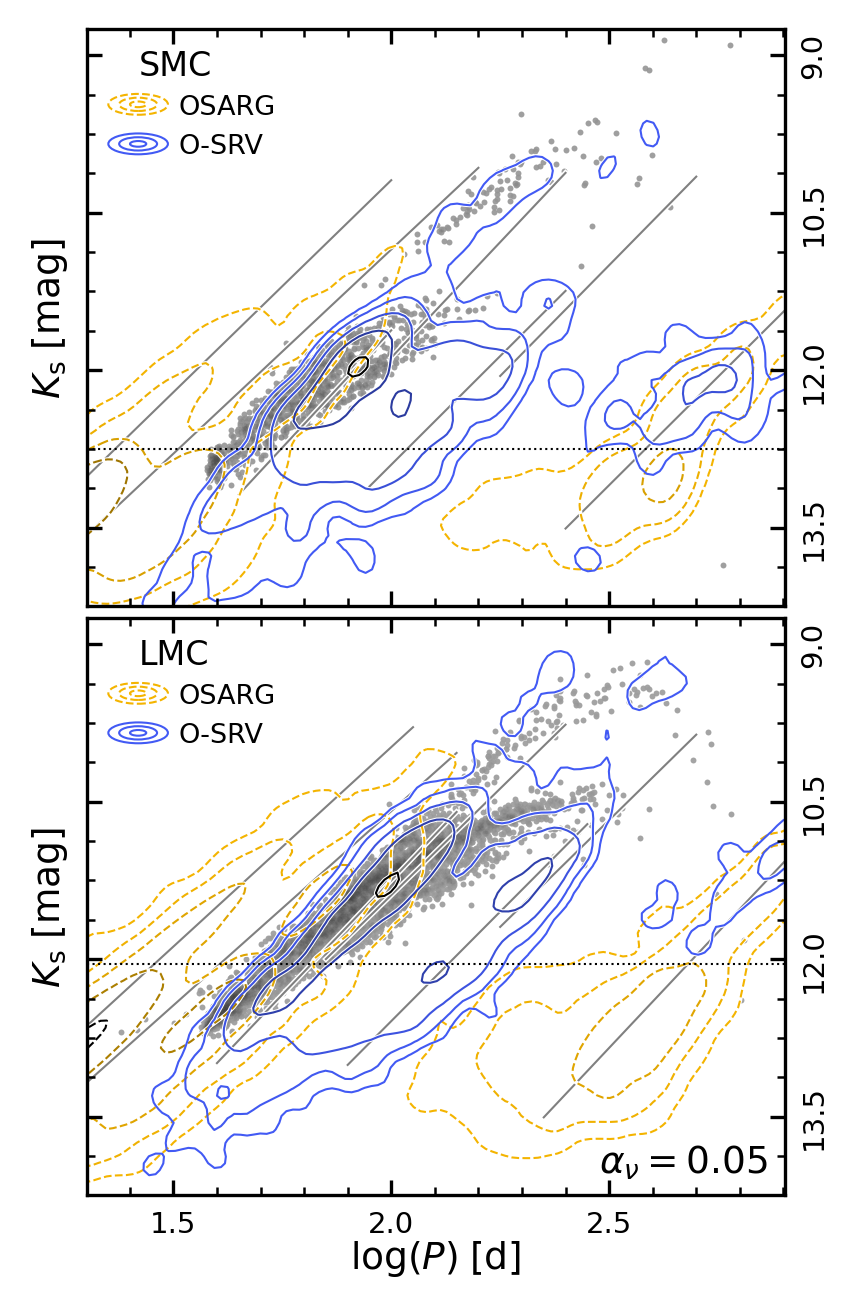}
    \includegraphics[width=0.33\textwidth]{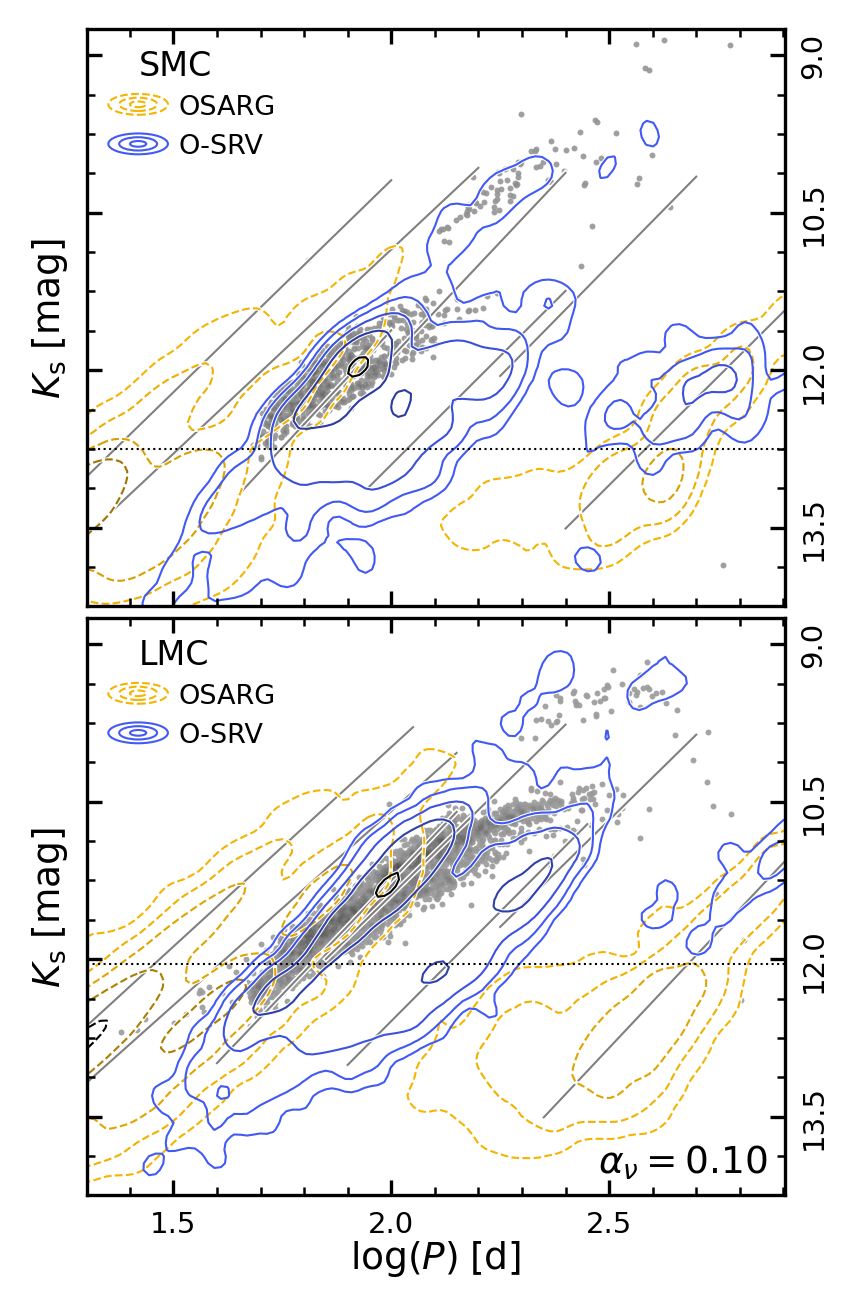}
    \includegraphics[width=0.33\textwidth]{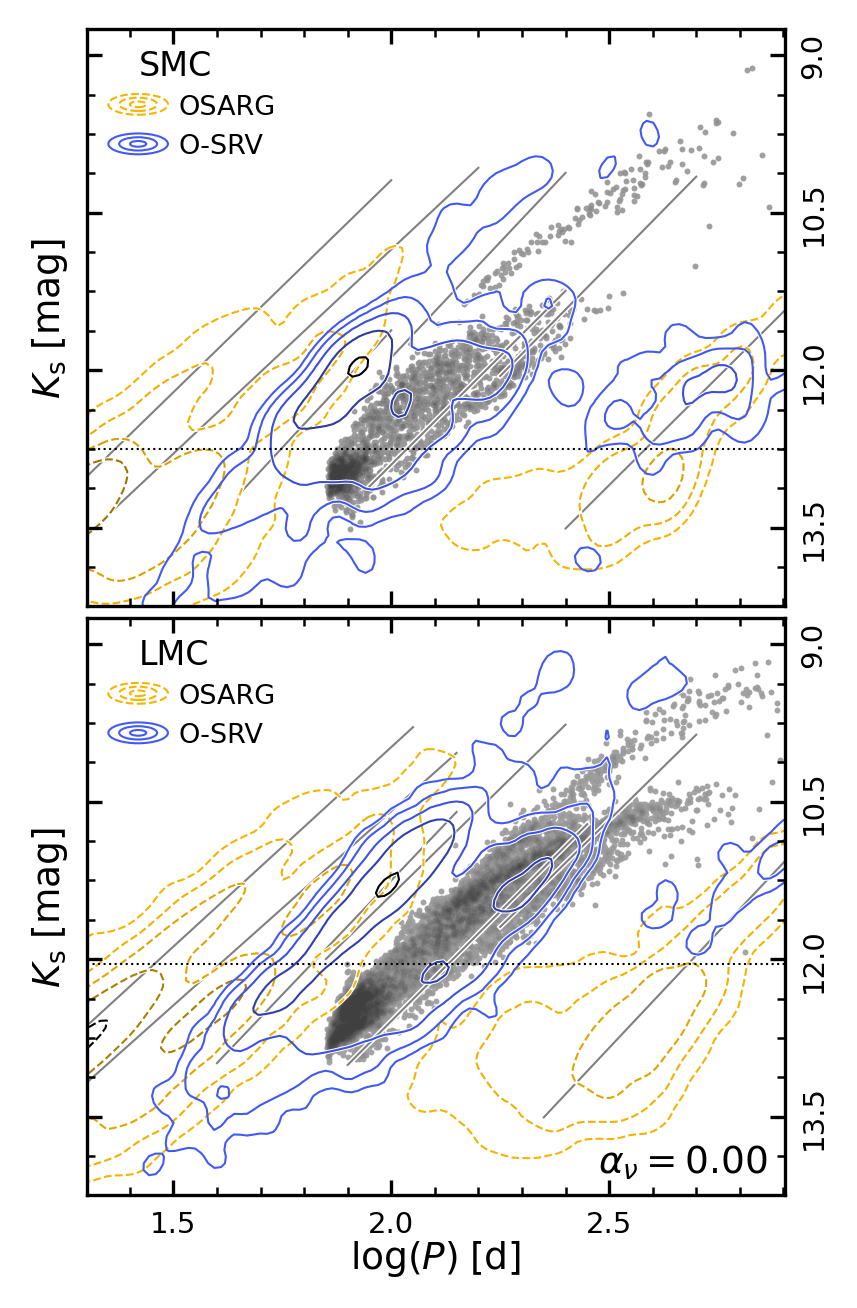}
    \includegraphics[width=0.33\textwidth]{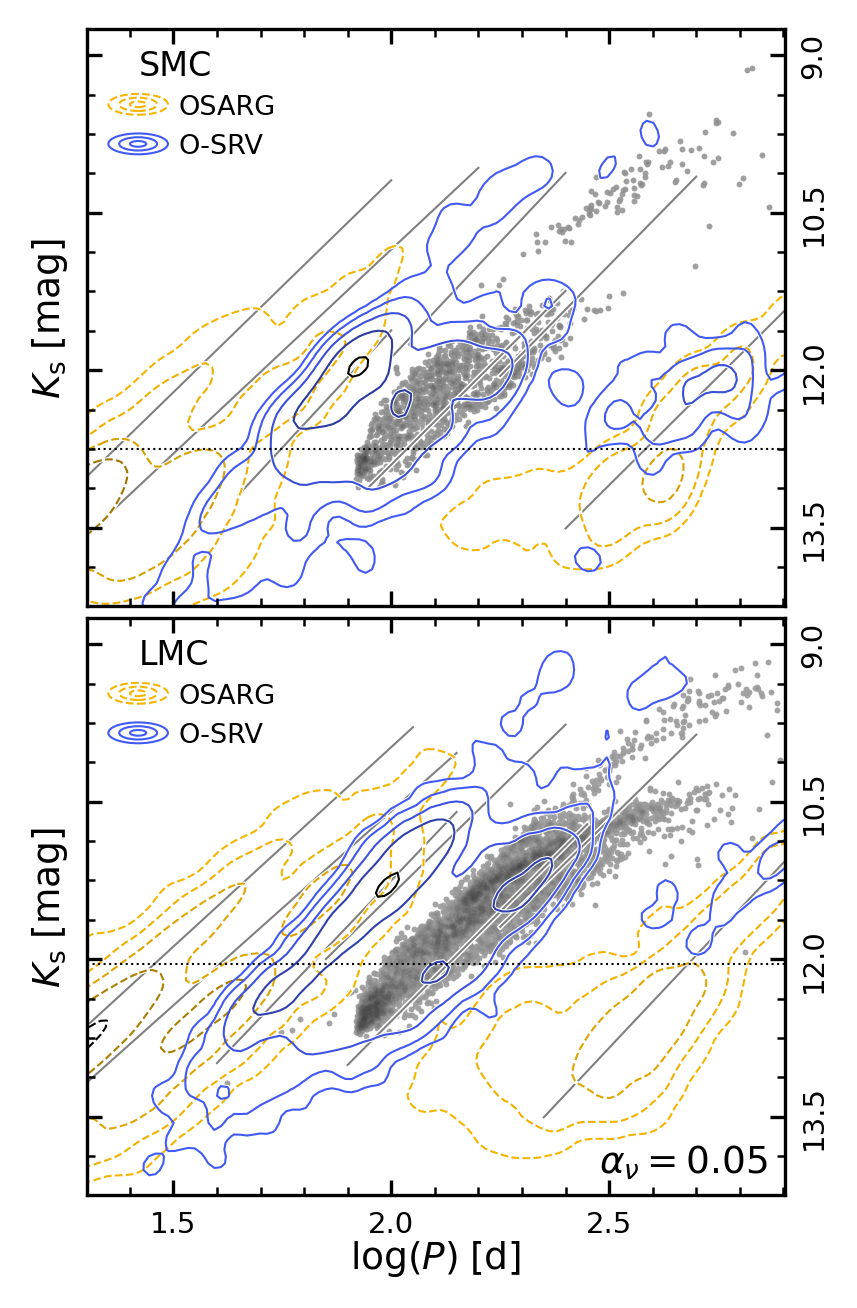}
    \includegraphics[width=0.33\textwidth]{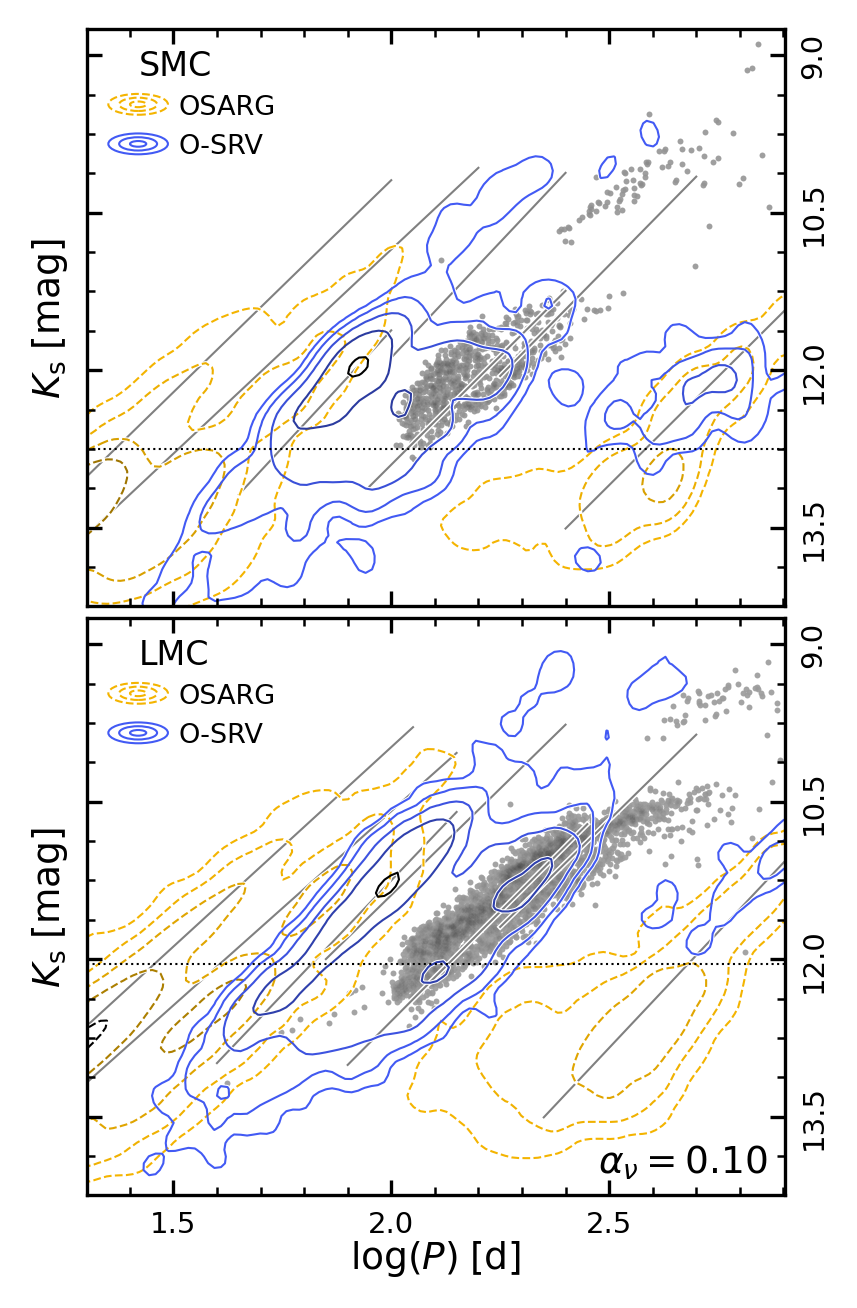}
    \caption{Similar to Fig.~\ref{fig:PLD_grid}, but limited to O-rich SRVs and comparing observations with synthetic stellar population models of the MCs. Gray dots correspond to the predicted periods of simulated stars unstable to self-excited pulsation, assuming they all pulsate predominantly in the 1OM (top two rows) or in the FM (bottom two rows). Panels from left to right correspond to increasing values of the turbulent viscosity parameter $\alpha_{\nu}=0.0$, $0.05$, and $0.10$.}
    \label{fig:PLD_sim_AGB_dom_ctr_O}
\end{figure*}

\subsection{Combination with Stellar Population Models}
\label{ssec:CombinationWithStellarPopulationModels}

The interpretation of the comparison between models and observations is complicated by the dependence of the former on chemical composition and input physics. In order to overcome this difficulty, we take advantage of the same interpolation scheme outlined in Sect.~\ref{ssec:TheInstabilityStrip} to combine the predictions from hydrodynamic calculations with synthetic stellar population models of the MCs computed with the \texttt{TRILEGAL} code \citep{Girardi_etal_2005}. The details of these simulations are given in \citet{Pastorelli_etal_2019,Pastorelli_etal_2020}, while recent updates to \texttt{TRILEGAL} are presented in \citet{DalTio_etal_2022,Chen_etal_2023,Mazzi_etal_2024}. Following this method, we determine the critical luminosity corresponding to the onset of self-excited pulsation for any given simulated star. All stars fainter than that value are considered stable against pulsation and are excluded from further analysis. We only consider simulated stars evolving on the RGB or AGB, and disregard the core helium-burning ones representative of RSGs as they are not present in the OGLE-III catalogs (they are too bright in optical filters and thus saturate). By means of this approach we can can compare a realistic model of the LPV populations in the MCs with observations without the need to make any assumptions on the input parameters to the pulsation code, with the exception of the turbulent viscosity parameter $\alpha_{\nu}$.

Since we aim for a comparison in the PLD, we need to compute 1OM and FM periods for pulsating simulated stars. This cannot be done using only the models presented here, except for the stars that just became unstable to pulsation. Therefore, we adopt the prescriptions of \citet{Trabucchi_etal_2019,Trabucchi_etal_2021a} to compute 1OM and FM periods, respectively. These authors also provide prescriptions to assess which one of the two modes is dominant. Unfortunately, the former is based on linear models and the latter disregards the effect of turbulent viscosity, so they are not applicable to the present study. In the absence of a better alternative, we are forced to adopt a pragmatic approach, and compare the results with observations under two simplified scenarios: in the first, we assume that all simulated stars pulsate predominantly in the 1OM, and in the second, we assume that they all pulsate in the FM. The main limitation of this approach is that we cannot study quantitatively the period distributions, and we can only perform a morphological comparison with the observed PLD. We stress that this has only a marginal impact on the conclusions of this study, as the main goal is to understand the location of the onset of instability in the PLD.

In the top two rows of Fig.~\ref{fig:PLD_sim_AGB_dom_ctr_O} we compare the distribution of simulated 1OM periods of O-rich stars with observations, and find a high degree of agreement between the model distribution and the morphology of the observed SRVs. In absence of turbulent viscosity, we find an excess of simulated stars for the SMC case, specifically near the RGB tip on the short-period side of sequence C$^{\prime}$, whereas the LMC case is well reproduced. If $\alpha_{\nu}$ is increased (central and right-hand columns of Fig.~\ref{fig:PLD_sim_AGB_dom_ctr_O}), thereby delaying the onset of pulsation, some simulated stars disappear at short periods, improving the comparison for the SMC while making it slightly worse in the LMC. We also note that stronger turbulent viscosity efficiently suppresses pulsation at bright magnitudes where relatively hot and massive stars are observed. These results suggest that a single value of $\alpha_{\nu}$ is not sufficient to reproduce the observed distribution of SRVs in both MCs (i.e. at different metallicity) and at all luminosities. Similar considerations apply to the FM periods shown in the bottom two rows of Fig.~\ref{fig:PLD_sim_AGB_dom_ctr_O}. We note that in both figures a ``plume'' at long periods and $10.5\lesssim K_{\rm s}/{\rm mag}\lesssim11.0$ is visible in the LMC simulation that has no counterpart in the observations. These are simulated stars that are expected to display large-amplitude FM pulsation, i.e. Mira-like variability, but they are not as prominent in the observations since most of the SMC Miras are C-rich.

\begin{figure*}
    \centering
    \includegraphics[width=0.4\textwidth]{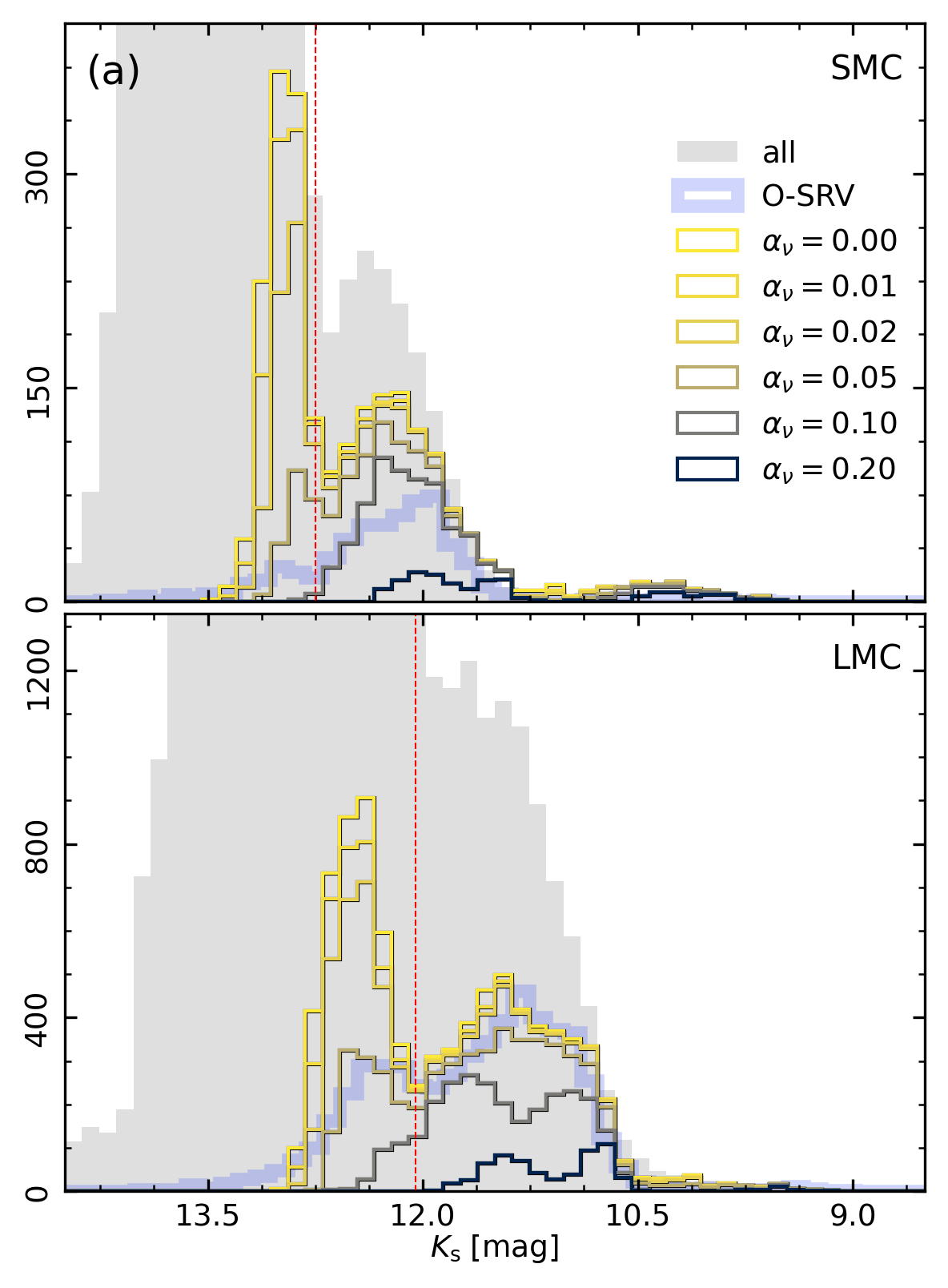}
    \includegraphics[width=0.4\textwidth]{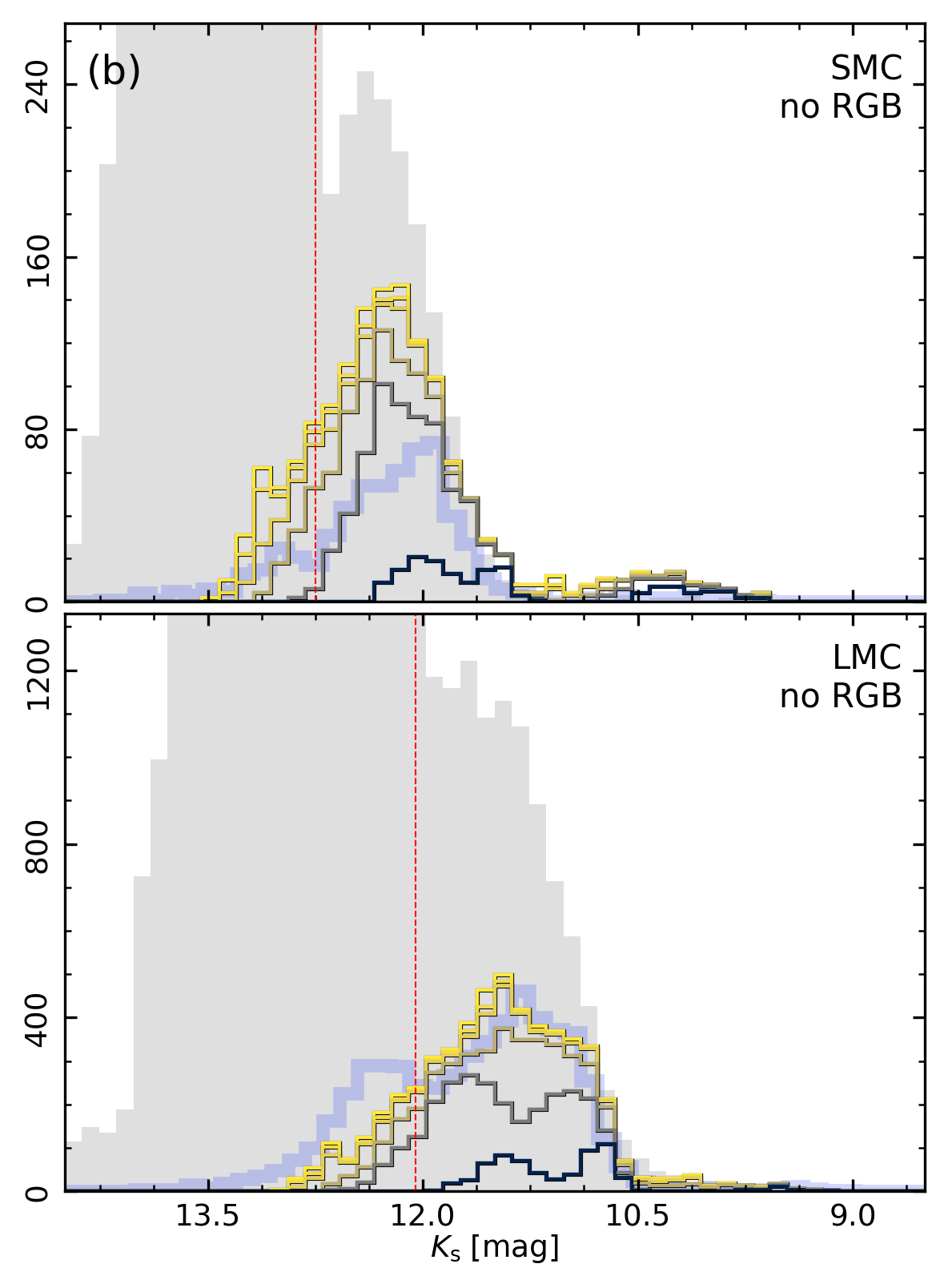}
    \includegraphics[width=0.4\textwidth]{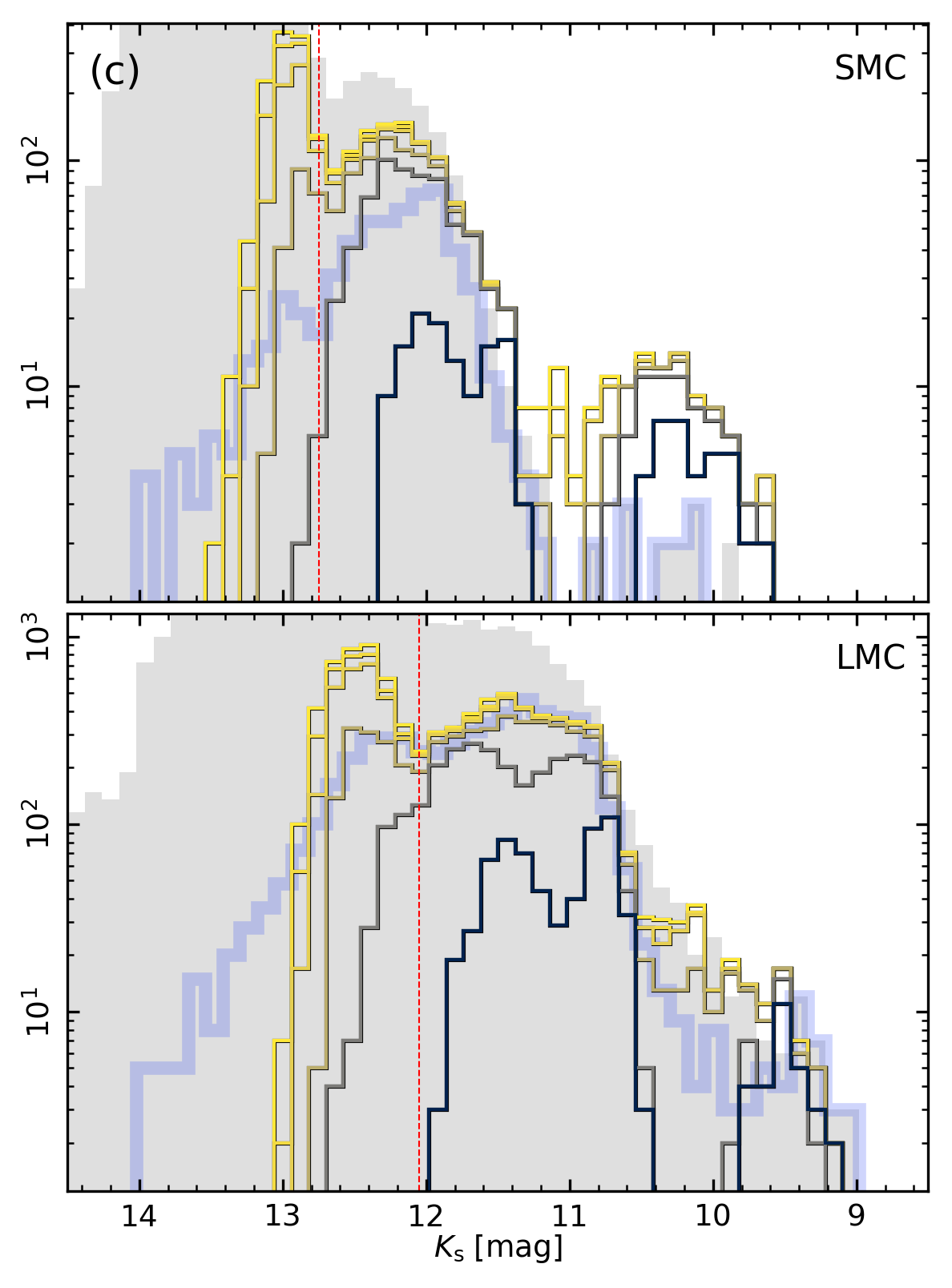}
    \includegraphics[width=0.4\textwidth]{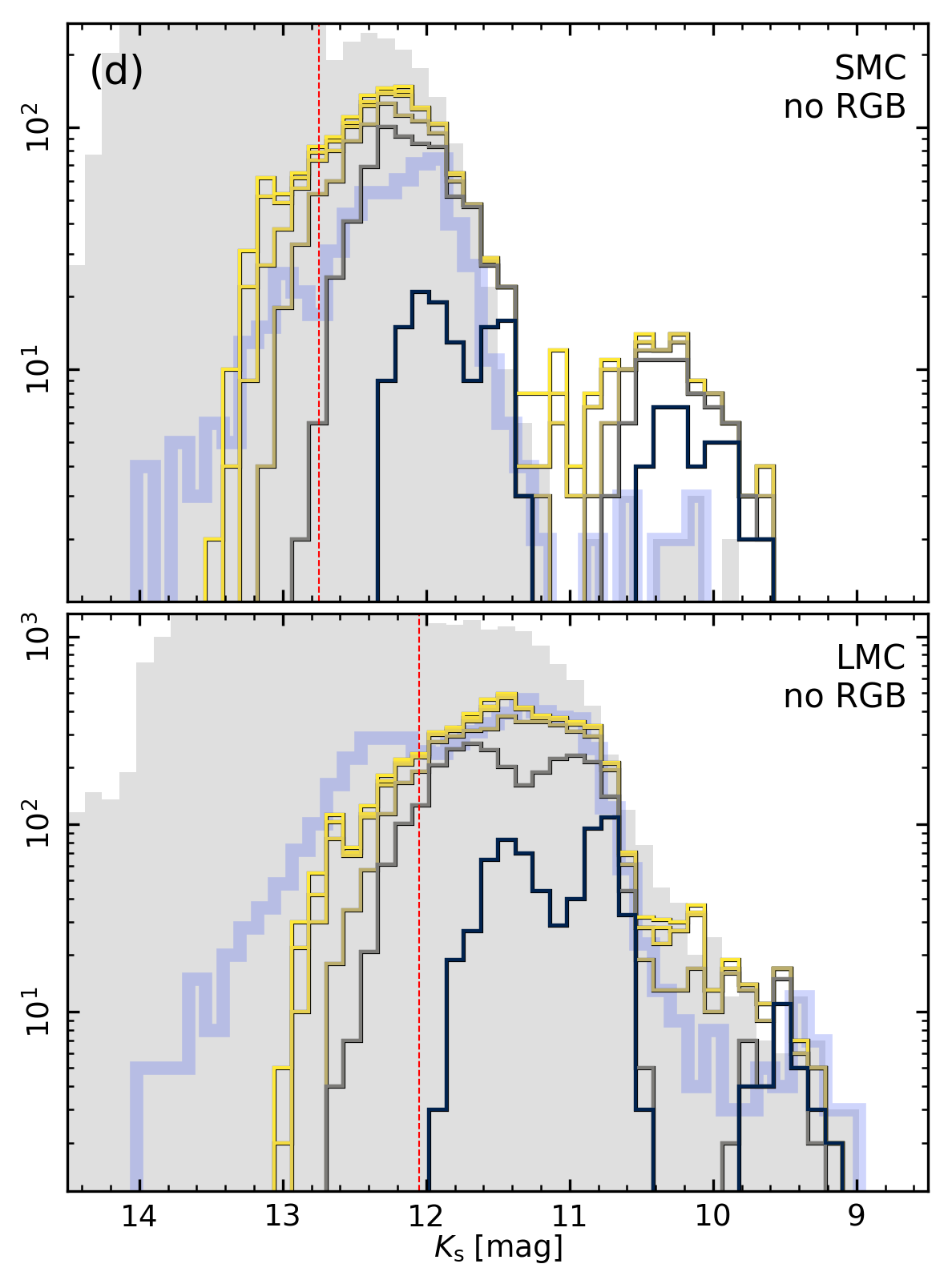}
    \caption{$K_{\rm s}$-band LFs of observed and simulated stars. The gray filled histogram shows the distribution of all observed stars, while the thick blue histogram shows the observed O-rich SRVs. Thin lines are the distributions of unstable stars from the simulation for various choices of $\alpha_{\nu}$, as indicated by the color-coding. Panels (a) and (c) were obtained including both RGB and AGB stars from the simulation, whereas RGB stars were excluded in producing the histograms in panels (b) and (d). The top (bottom) part of each panel corresponds to the SMC (LMC), as indicated. Panels (c) and (d) are the same as panels (a) and (b), respectively, except that the vertical axis is in logarithmic scale. LSPs are omitted from all observed distributions.}
    \label{fig:hist_Ks}
\end{figure*}

Many simulated stars near the end of their RGB phase are predicted to be unstable to self-excited pulsation, as can be seen by the overdensity just below the RGB tip in Fig.~\ref{fig:PLD_sim_AGB_dom_ctr_O}. Based on their predicted periods, these seem to be in better agreement with the observed OSARGs than with the SRVs, especially in the SMC. In order to examine this aspect in more detail, we opt to compare models and observations in terms of their $K_{\rm s}$-band luminosity functions (LFs) in Fig.~\ref{fig:hist_Ks}. Besides being more quantitative, this approach is preferable to the analysis of the PLD. Indeed, by construction, the LFs displayed by the synthetic population models are in agreement with observations as the models are calibrated to reproduce them \citep[see][]{Pastorelli_etal_2019}. In contrast, pulsation predictions are still uncalibrated and bound to show a higher degree of uncertainty. The observed LFs of O-rich SRVs displayed in Fig.~\ref{fig:hist_Ks} clearly show a bump associated with the RGB tip, indicating that SRVs include a significant fraction of stars that have not yet reached the AGB phase. Low values of $\alpha_{\nu}$ lead to a significant overestimate of the number of RGB stars unstable to pulsation, yet they are unable to account for the much higher number counts of OSARGs. At the same time, if the turbulent viscosity is too large the models do not reproduce the observed LFs of SRVs. 

The bottom part of panel (a) in Fig.~\ref{fig:hist_Ks} compares the observed and predicted LFs in the LMC. One can see that $\alpha_{\nu}$ cannot be larger than about 0.02 in order to reproduce the observed distribution above the tip of the RGB. However, with such a value of $\alpha_{\nu}$, the simulation predicts about twice the observed O-rich SRVs below the RGB tip. In other words, a lower turbulent viscosity parameter should be adopted for AGB models compared to RGB ones. This suggests that turbulent viscosity is not constant, but rather depends on the stellar properties or evolutionary phase. This indication is further supported by the fact that such values would not be appropriate for stars in the SMC, shown in the top part of panel (a) in Fig.~\ref{fig:hist_Ks}. Indeed, in order to reproduce the observed LFs in the SMC, one would need to adopt a value $0.05\lesssim\alpha_{\nu}\lesssim0.1$ below the RGB tip and $0.1\lesssim\alpha_{\nu}\lesssim0.2$ above it. Therefore, as opposed to the LMC case, the turbulent viscosity parameter must be lower in RGB models than in AGB models.

\section{Discussion}
\label{sec:Discussion}

\subsection{The Driving Mechanism of LPVs}
\label{ssec:TheDrivingMechanismOfLPVs}
The analysis of our results shows that the onset of instability is sensitive to the strength of superadiabatic convection in the subphotospheric layers, as well as on the hydrogen mass fraction in the envelope. In view of the good agreement with observations, this represents a strong indication that the zones of partial ionization of major elements play a primary role in the excitation of pulsations, reinforcing the idea that the driving mechanism of SRVs and Miras is akin to the heat mechanism of classical pulsators \citep[as already suggested by][]{Ostlie_Cox_1986} or to a $\gamma$-type mechanism \citep[probably dominated by convection, as proposed by][]{Munteanu_etal_2005}. Metallicity effects fit this picture through their impact on the opacity and thus on the thermal structure of the evelope, which in turn affects the depth of partial ionization zones. As a result, the IS becomes redder for higher metal content, as is the case for core helium-burning pulsators \cite[e.g.][]{DeSomma_etal_2022}.

Hydrodynamic pulsation models provide a means to investigate how the envelopes of evolved stars develop self-excited pulsation, as opposed to other driving mechanisms such as stochastic excitation. Our results indicate that pulsation emerges in the proximity of the transition from OSARGs to SRVs in the PLD, with the majority of OSARGs being intrinsically stable to pulsation, and the majority of SRVs being unstable. This supports the hypothesis that OSARGs are stochastically excited \citep{Soszynski_etal_2004}, and strongly suggests that the very difference between these two variability types lies in the underlying excitation mechanism. Evidence supporting this interpretation stems from the fact that none of the models we computed develops stable cycle pulsations dominated by any overtone mode higher than the first. This is consistent with studies that specifically linked SRVs with dominant pulsation in the 1OM or in the FM \citep{Hartig_etal_2014,Trabucchi_etal_2017,Trabucchi_etal_2021b,Yu_etal_2020}.

Yet, it cannot be excluded a priori that at least some OSARGs undergo self-excited pulsations. Indeed, OSARGs are known to display dominant 1OM pulsation on sequence B, in compatibility with some of our models. By itself, our grid of hydrodynamic models cannot provide a definitive answer to this question, and it needs to be coupled with a more realistic description of the populations of evolved stars in the MCs, as presented in Sect.~\ref{ssec:CombinationWithStellarPopulationModels}. This approach shows that the distribution of simulated AGBs unstable to pulsation match remarkably well the SRV sequences C$^{\prime}$ and C, respectively, depending on whether the models are assumed to pulsate in the 1OM or FM, and on the choice of the free parameter $\alpha_{\nu}$. However, this method also predicts the existence of a small population of self-excited RGB stars within about 0.2 mag from the RGB tip. On the one hand, these stars are necessary to explain the observed $K_{\rm s}$-band LF of SRVs. On the other hand, they have linear 1OM periods in the range of 30 to 45 days, which are too short for SRVs and rather end up on the OSARG B sequence once again.

This discrepancy is not easily resolved, especially given the sensitivity of the models to the uncertain turbulent viscosity. Nonetheless, the notion that the majority of OSARGs undergo stochastically driven oscillations and only a small fraction of them display self-excited pulsations seems unlikely. Indeed, if this were the case, one would expect a noticeable change in variability behavior between stochastically driven and self-excited OSARGs near the RGB tip. The fact that this is not observed is difficult to reconcile with the clean association between the clear-cut transition from OSARGs to SRVs above the RGB tip and the clear onset of self-excited pulsation in the models. It is true that stars near the RGB tip are physically very similar to each other, so that one could imagine a very smooth transition from stochastically driven to self-excited pulsation, but a comparable likeness is found above the RGB tip, so this explanation is not satisfactory. Therefore, it seems that the correct interpretation is the simplest one: SRVs are self-excited pulsators, and OSARGs are not. In this scenario, models can be reconciled with observations by adopting the appropriate value of the turbulent viscosity parameter $\alpha_{\nu}$. For instance, using $\alpha_{\nu}\gtrsim0.05$ for the LMC simulation correctly reproduces the LFs below the RGB tip, and results in a better agreement in the PLD (see Fig.~\ref{fig:PLD_sim_AGB_dom_ctr_O} and Fig.~\ref{fig:hist_Ks}). At the same time, doing so degrades the level of agreement at brighter magnitudes, both in the LF and in the PLD, consistent with the idea that the turbulent viscosity is a function of the stellar parameters.

\subsection{Implications}
\label{ssec:Implications}

\subsubsection{Calibration of Pulsation Models}
\label{sssec:CalibrationOfPulsationModels}

The impact of $\alpha_{\nu}$ on the PLD can be stated in other words by saying that, by adjusting its value, one can shape the short-period edge of the SRV sequence C$^{\prime}$. This provides a promising way to calibrate the turbulence viscosity, as an alternative to previously proposed methods. For instance, \citet{Olivier_Wood_2005} indicate that turbulent viscosity should be tuned so that models match the observed pulsation amplitudes, but the latter can be affected by atmospheric dynamics, adding a layer of complexity that is difficult to control. The framework introduced by \citet{Trabucchi_etal_2017} and applied in the present work represents a promising tool to characterize the PLD of LPVs on theoretical grounds, provided the role of turbulent viscosity is properly constrained. Such a calibration would substantially enhance the predictive power of pulsation models, but it requires some caution. Indeed, it can be applied to improve the agreement with observations by effectively removing stars from the left-hand side of the sequence, but it cannot add any to the right-hand side. This becomes critical when dealing with the long-period edge of the C$^{\prime}$ sequence, as well as the transition to sequence C. At the current stage we are unable to address this question as we lack the appropriate criterion to characterize the 1OM-FM switch. However, it is worth noticing that simulated PL sequences appear to be less steep than the observed ones, i.e. they lean toward longer periods at bright magnitudes (see Figure~1 of \citet{Trabucchi_etal_2017}) and towards shorter periods at faint magnitudes, more than is observed in the MCs. This mismatch cannot be mitigated through the choice of $\alpha_{\nu}$.

Since pulsation periods are very sensitive to the temperature of the envelope, this trend may rather be traced to the adopted temperature scale. In this sense, it is relevant to recall that the synthetic population models adopted here are known to display a slight, systematic color excess with respect to observations, a second-order effect with negligible impact on the LFs and on the overall calibration of the evolutionary models. This behavior has been investigated by \citet{Pastorelli_etal_2019}, who excluded opacity and composition effects as its cause, but concluded that it can be explained by a temperature offset. This aspect cannot be overlooked when calibrating pulsation models, and it is rather advisable to perform a joint calibration of the temperature scale and the turbulent viscosity parameter, taking advantage of both photometric and variability observations for this purpose.

\subsubsection{Distance and age determinations}
\label{sssec:DistanceAndAgeDeterminations}

The potential of LPVs as standard candles is widely recognized, and while Miras have been the focus of most studies in this sense \citep[see][for a recent discussion]{Huang_2024}, SRVs have been recently acknowledged as promising candidates toward this purpose \citep{Rau_etal_2019,Trabucchi_etal_2021b,Hey_etal_2023}. Due to the intimate correlation between stellar luminosity, mass, and main-sequence lifetime, LPVs carry also interesting potential as age indicators \citep{Catchpole_etal_2016,Grady_etal_2019,Sanders_etal_2022,Sanders_etal_2024,Trabucchi_Mowlavi_2022,Trabucchi_2024}. Crucially, these applications implicitly rely on the assumption that the PL relations of LPVs (especially of Miras) are to a large degree independent of chemical composition. This view has recently been challenged on observational grounds \citep[][and references therein]{Chen_etal_2024}, although no conclusive evidence has been presented yet. Moreover, \citet{Trabucchi_etal_2021b} have drawn attention to the fact that photometric amplitudes of variability, customarily used to select dominant periods and discriminate between SRVs and Miras (even for the purpose of choosing candidate distance indicators), are likely to be sensitive to the envelope metal content as they are controlled by the spectral absorption features of O- and C-bearing molecules.

Various studies based on linear pulsation models \citep{Fox_Wood_1982,Wood_etal_1983,Ostlie_Cox_1986,Bedijn_1988,Groenewegen_DeJong_1994} suggest that, at fixed mass, luminosity, and temperature, periods are independent of chemical composition. The analysis by \citet{Trabucchi_etal_2019} indicates that any effect of metallicity, hydrogen abundance, or carbon-to-oxygen ratio is indirect and operates through the effect of radiative opacity on the surface temperature and radius. However, many of the aforementioned studies include some form of composition dependence in their prescription for the onset of pulsation or shift between pulsation modes. In this work, we find clear evidence that the stability against pulsation is sensitive to the chemical composition of the envelope. Even more importantly, we find that different values of the turbulent viscosity parameter are needed to reproduce the observations in the LMC and SMC, suggesting a much more complex role of composition in the pulsation properties of LPVs than previously thought. A systematic investigation of these aspects, based on both observations and models, is therefore required.

\subsubsection{Mass loss and long secondary periods}
\label{sssec:MassLossAndLongSecondaryPeriods}

According to our results, relatively faint and hot LPVs should be intrinsically stable to pulsation, so that their OSARG-type variability can be attributed to solar-like oscillations in a few low-order, radial and nonradial modes. As they evolve on the RGB or AGB, these stars enter the IS, become unstable to self-excited pulsation, and become SRVs.

This scenario fits nicely in the interpretation of the PLD which has emerged in the recent years. \citet{Trabucchi_etal_2017} argued that sequence B OSARGs are 1OM-dominated stars exactly as are SRVs on sequence C$^{\prime}$ \citep[a modal assignment confirmed by][]{Yu_etal_2020}, and that the gap between these sequences is an observational bias resulting from LSPs not being properly accounted for. \citet{McDonald_Trabucchi_2019} linked the transition between these sequences with a step-up in the mass-loss rate, consistent with previous evidence \citep[e.g.][]{Nicholls_etal_2009,Wood_Nicholls_2009} relating LSPs with mass-ejection episodes and circumstellar dust. More recently, \citet{Soszynski_etal_2021} \citep[see also][]{Pawlak_etal_2024,Goldberg_etal_2024} provided very convincing evidence of the binary origin of LPSs, which requires the presence of circumstellar material.

Further investigation is needed to characterize the role of the onset of self-excited pulsation in this scenario, but it is hard to believe that the association of this instability with the transition from sequence B to sequence C$^{\prime}$ is a coincidence. In particular, we note that stochastic driving is characterized by a rather inefficient input of energy into oscillations \citep[see, e.g., the analysis by][]{Buchler_etal_2004}, while self-excited pulsations entail a significantly more coherent atmospheric dynamics known to play a key role in dust-driven stellar winds \citep{Hoefner_Olofsson_2018}.

According to \citet{Winters_etal_2000}, pulsation can drive moderate mass loss even when dust is scarce and radiation pressure is ineffective. We can thus speculate that the onset of self-excited pulsation at the transition between sequences B and C$^{\prime}$ is at least partially responsible for the corresponding increase in mass-loss rate reported by \citet{McDonald_Trabucchi_2019}, and that it can provide the necessary conditions for the occurrence of LSPs.
%We can thus outline the following scenario. Relatively faint and hot LPVs are intrinsically stable to pulsation, but undergo solar-like oscillations in a few low-order radial and non-radial modes as OSARGs. As they evolve on the RGB or AGB, they enter the instability strip and they become unstable to self-excited pulsation and become SRVs. While stochastic driving is characterized by a rather inefficient input of energy into oscillations \citep[see e.g. the analysis by][]{Buchler_etal_2004}, self-excited pulsations entail a significantly more coherent atmospheric dynamics, capable of driving moderately intense mass loss \citep{Hoefner_Olofsson_2018}.

Once gas and dust start to be injected into the circumstellar environment, an otherwise unseen (sub)stellar companion can become detectable by virtue of its interaction with them, for instance modulating the dust in its orbit so that it periodically obscures the primary star \citep[or reveals it; see][]{Goldberg_etal_2024}. Hydrodynamic atmosphere models properly coupled with interior pulsation simulations are necessary to verify this hypothesis. If confirmed, this would provide a further link between the complex processes experienced during the evolution on AGB, and a further way to exploit observations for constraining stellar models.

\section{Conclusions}
\label{sec:Conclusions}

We perform the first systematic investigation of the onset of self-excited radial pulsation in LPVs using one-dimensional hydrodynamic simulations of red giant envelopes. Our calculations cover a substantial portion of the space of stellar parameters in terms of mass, luminosity, temperature, and chemical composition, representative of the AGB and of the brightest portion of RGB. We account for various choices of the turbulent viscosity parameter playing a critical role in the emergence of pulsation. By combining the results with synthetic stellar population models of the MCs, we can directly compare the predictions of our models with observations in the PLD.

Using this approach, we are able to firmly establish a connection between the onset of nonlinear instability and the transition from low-amplitude red giants to SRVs, more precisely from OSARGs to SRVs in the OGLE classification. While SRVs undergo self-excited pulsations, OSARGs do not. This result agrees with previous theoretical \citep{Xiong_etal_2018,Cunha_etal_2020} and observational evidence \citep{Banyai_etal_2013} supporting a transition in driving mechanism, and confirm the idea that OSARGs undergo stochastically driven solar-like oscillations \citep{Soszynski_etal_2004}. Furthermore, we lend additional support for the binary explanation of LSPs in LPVs by providing a direct, natural explanation for the increase in mass-loss rates associated with transition between sequences B and C$^{\prime}$. The resulting injection of dust in the circumstellar environment is believed to be a key ingredient to explain LSPs.

We examine the onset of pulsation in terms of the IS of SRVs. We find that its blue edge is approximately vertical in the HRD, and it moves redward with increasing metallicity, hydrogen content, and turbulent viscosity, and with decreased efficiency of superadiabatic convection. This behavior is comparable with that of the IS of core helium-burning stars, and tends to confirm the important role played by the partial hydrogen ioniziation zone advocated by previous studies \citep{Ostlie_Cox_1986}. Therefore, our research supports the hypothesis that self-excited pulsations in luminous red giants are powered by a mechanism similar to the heat engine of classical pulsators.

The turbulent viscosity remains a critically uncertain ingredient in our models. We find that a nonconstant value is required to reproduce the location of the PL sequence of SRVs in the MCs, as well as their $K_{\rm s}$-band LFs. In particular, we find that turbulent viscosity should decrease with metallicity to achieve a good match with observations in both the MCs. In order to increase the predictive power of pulsation models, a calibration of the turbulent viscosity parameter as a function of stellar parameters is highly desirable. Given the high sensitivity of pulsation periods on the surface temperature, it is highly advisable that this procedure be performed in conjunction with a calibration of the temperature scale of stellar models. Such a calibration is beyond the scope of this paper, but we provide a solid basis for future work in this direction. This is especially true in view of the extensive variability data from the \textit{Gaia} mission \citep{Lebzelter_etal_2023,GaiaCollaboration_Trabucchi_etal_2023}, which already proved powerful for studying LPVs \citep[e.g.][]{Lebzelter_etal_2018,Lebzelter_etal_2019}, and from the Legacy Survey of Space and Time that will operate at the upcoming Vera C. Rubin Observatory \citep{Ivezic_etal_2019}.

Finally, our models add further support to the potentially problematic fact that the PL relation of LPVs might be sensitive to chemical composition. If confirmed, this could severely impact the potential of LPVs as standard candles and age indicators, unless properly accounted for. Therefore, a systematic theoretical and observational investigation of the chemical dependence of LPV pulsation is strongly recommended. Table~\ref{tab:results} in machine-readable form and the accompaning interpolation routine are made publicly available via\dataset[DOI: 10.5281/zenodo.14002294]{https://doi.org/10.5281/zenodo.14002294} \citep{14002294}. The hydrodynamic time series presented in this work will be made available upon request to the author.

\vspace{2.5cm}
\noindent We wish to express our deepest gratitude to our mentor Paola Marigo for her unwavering support and encouragement in both our academic and personal lives. Her teachings will always remain with us and guide us as we shape our \mbox{own paths.}

\begin{acknowledgements}
We thank the anonymous referee for the useful comments and suggestions that to improve this paper. We acknowledge financial support from the Supporting TAlent in ReSearch@University of Padova (STARS@UNIPD) for the project ``CONVERGENCE: CONstraining the Variability of Evolved Red Giants for ENhancing the Comprehension of Exoplanets.'' M.T. acknowledges the support provided by the Swiss National Science Foundation through grant No. 188697. This research has made use of data from the \mbox{OGLE-III} Catalogue of Variable Stars, data products from the Two Micron All Sky Survey \citep[2MASS;][]{doi2mass}, which is a joint project of the University of Massachusetts and the Infrared Processing and Analysis Center/California Institute of Technology, funded by the National Aeronautics and Space Administration and the National Science Foundation. This research has made use of NASA's Astrophysics Data System Bibliographic Services, and of the following services provided by CDS, Strasbourg: the SIMBAD database, the VizieR catalog access tool \citep[DOI:10.26093/cds/vizier;][]{vizier}, the ``Aladin sky atlas'' \citep{aladin}, and the cross-match service \citep{cds2012,cds2020}.
\end{acknowledgements}

\software{
\texttt{COLIBRI} \citep{Marigo_etal_2013}; \texttt{TRILEGAL} \citep{Girardi_etal_2005}; \texttt{{\AE}SOPUS} \citep{Marigo_etal_2022,Marigo_etal_2024a,Marigo_etal_2024b}; \texttt{OP} \citep{Seaton_2005}; Starlink Tables Infrastructure Library \citep[\texttt{STILTS} and \texttt{Topcat};][]{stilts}; \texttt{IPython} \citep{ipython}; \texttt{Jupyter} \citep{jupyter}; \texttt{NumPy} \citep{numpy}; \texttt{SciPy} \citep{scipy}; \texttt{matplotlib} \citep{matplotlib}; and \texttt{Astropy} \citep{astropy2013,astropy2018,astropy2022}.
}

\bibliographystyle{aasjournal}
\bibliography{references.bib}

\end{document}